\documentclass[onecolumn,draft]{IEEEtran}
 \IEEEoverridecommandlockouts

\usepackage{amsthm,amssymb,bbm,bm,subfigure,color,caption,euscript,tikz,pgfplots,tkz-berge}
\theoremstyle{plain}      
\newtheorem{lemma}{Lemma} 
\newtheorem{corollary}{Corollary}
\newtheorem{theorem}{Theorem}

\newtheorem{exam}{Example}

\theoremstyle{definition}
\newtheorem{definition}{Definition}

\theoremstyle{remark}
\newtheorem{claim}{Claim}
\newtheorem{remark}{Remark}

\usepackage{graphicx}
\graphicspath{{Figures/}} 
\usepackage[cmex10]{amsmath}
\newcommand{\defeq}{\overset{\text{def}}{=}}

\begin{document}

\title{Distributed Function Computation \\ Over a Rooted Directed Tree}

\author{\IEEEauthorblockN{Milad Sefidgaran and Aslan Tchamkerten}

\IEEEauthorblockA{Department of Communications and Electronics\\
Telecom ParisTech\\
 \{sefidgaran, aslan.tchamkerten\}@telecom-paristech.fr}
 
 \thanks{This paper was presented in part at ITW 2013.}
 \thanks{This work was supported in part by a ``Future et Rupture'' grant from
the Institut Telecom, and by an Excellence Chair Grant
 from the French National Research Agency (ACE project).}
}
\maketitle

\begin{abstract}

%\boldmath

This paper establishes the rate region for a class of source coding function computation setups where sources of information are available at the nodes of a tree and where a function of these sources must be computed at the root. The rate region holds for any function as long as the sources' joint distribution satisfies a certain Markov criterion. This criterion is met,  in particular, when the sources are independent.

This result recovers the rate regions of several function computation setups. These include the point-to-point communication setting with arbitrary sources, the noiseless multiple access network with ``conditionally independent sources,'' and the cascade network with Markovian sources.

\end{abstract}
\section{Introduction}  
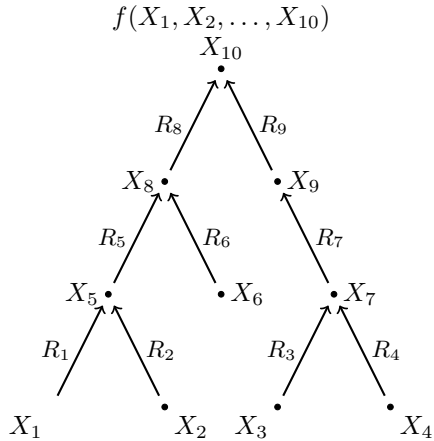
\begin{figure}
\begin{center}
\begin{tikzpicture}[scale=0.75]
r\draw [black,fill=black] (2,0) circle [radius=0.05];
\draw [black,fill=black] (4,0) circle [radius=0.05];
\draw [black,fill=black] (6,0) circle [radius=0.05];
\draw [black,fill=black] (3,2) circle [radius=0.05];
\draw [black,fill=black] (5,2) circle [radius=0.05];
\draw [black,fill=black] (4,4) circle [radius=0.05];
\draw [black,fill=black] (3,6) circle [radius=0.05];
\draw [black,fill=black] (1,2) circle [radius=0.05];

\draw [->][thick](4.1,0.2)--(4.9,1.8);
\draw [->][thick](0.1,0.2)--(0.9,1.8);
\draw [->][thick](1.9,0.2)--(1.1,1.8);
\draw [->][thick](3.9,4.2)--(3.1,5.8);
\draw [->][thick](4.9,2.2)--(4.1,3.8);
\draw [->][thick](5.9,0.2)--(5.1,1.8);
\node [below left] at (0,0) {$X_1$};
\node [below right] at (2,0) {$X_2$};
\node [below left] at (4,0) {$X_3$};
\node [below right] at (6,0) {$X_{4}$};
\node [left] at (1,2) {$X_5$};
\node [right] at (3,2) {$X_6$};
\node [right] at (5,2) {$X_7$};
\node [right] at (4,4) {$X_9$};
\node [above] at (3,6) {$X_{10}$};
\node [above] at (3,6.5) {$f(X_1,X_2,\ldots,X_{10})$};
\node [left] at (2,4) {$X_8$};

\node [left] at (0.5,1) {\small{$R_1$}};
\node [right] at (1.5,1) {\small{$R_2$}};
\node [left] at (4.5,1) {\small{$R_3$}};
\node [right] at (5.5,1) {\small{$R_{4}$}};
\node [left] at (1.5,3) {\small{$R_5$}};
\node [right] at (2.5,3) {\small{$R_6$}};
\node [right] at (4.5,3) {\small{$R_7$}};
\node [left] at (2.5,5) {\small{$R_8$}};
\node [right] at (3.5,5) {\small{$R_9$}};

\draw [->][thick](1.1,2.2)--(1.9,3.8);
\draw [->][thick](2.9,2.2)--(2.1,3.8);
\draw [black,fill=black] (2,4) circle [radius=0.05];
\draw [->][thick](2.1,4.2)--(2.9,5.8);

\end{tikzpicture}
\end{center}
\caption{Distributed computation on a rooted directed tree.}
\label{fig:rootedDirectedTree2}
\end{figure}

Consider a directed tree network with $k\geq 1$ nodes where each edge points towards the root.  An example of such a network is depicted in Fig.~\ref{fig:rootedDirectedTree2}. Source $X_u$, $u\in \{1,2,\ldots,k\}$, is available at vertex $u$ and a given function $f(X_1,X_2,\ldots,X_{k})$ must be computed at the root. Communication occurs in multiple hops, losselessly, from level one (composed of sources $X_1,X_2,X_3,X_4$ in the example) up to the root where the function is finally computed.  Tree networks generalize some previously investigated settings including point-to-point \cite{OrliRoc01}, multiple access \cite{KornMar79,SefiTchISIT11},\footnote{By multiple access we intend a noiseless multiple access channel where the receiver gets separate streams of data from each of the sources.} and cascade (relay-assisted) \cite{CuffSuElg09,Visw10,Shay11,SefiTch122}, and can be used as backbones for computing functions over general networks \cite{KowsKum12}.

Given a tree, a function $f$, and a joint distribution over the sources $(X_1,X_2,\ldots,X_{k})$ we seek to characterize the least amounts of information that need to flow across the tree edges so that the function can be computed with arbitrarily high probability in the limit of multiple i.i.d. instances of the sources. 
In this paper, we first provide a cut-set outer bound to the rate region which generalizes the outer bounds established in \cite[Corollary~2]{SefiTchISIT11} for the multiple access network and in \cite[Theorem~2]{Visw10} for the cascade network. Second, we establish an inner bound to the rate region which generalizes the  inner bound for multiple access derived in \cite[Proposition~1]{SefiTchISIT11}. We then derive the main result which gives a sufficient condition on the sources' joint distribution under which the inner and outer bounds are equal. This condition is satisfied, in particular, when the sources are independent. Through this result we recover all the previously known rate regions related to network configurations without interaction.\footnote{By interaction we mean a network configuration that contains a pair of sources for which information can flow both ways. The simplest example is the two-way-two-node case.} These include point-to-point communication, multiple access, and  cascade network configurations---see Theorem~\ref{th:treeRate} thereafter.

\subsection*{Related Works}
Communication in distributed function computation has been investigated both under the zero error probability criterion and the asymptotic zero error probability criterion.\footnote{The problem considered here should be distinguished from gossip algorithms \cite{Shah09}, distributed decision making \cite{tsit84}, belief propagation algorithms \cite{Pear82}, and population protocols \cite{AspnRup07}, where the goal is to compute one or multiple (deterministic or probabilistic) functions at all the nodes.}  We review related works separately as these criteria, although conceptually similar, can yield very different results and often involve different analysis---zero error problems are typically more combinatorial.
\subsubsection*{\sc Zero-error probability}
Computational complexity has traditionally been measured in terms of the number of primitive operations required to compute a given function. When computation is carried out in a distributed fashion, Abelson \cite{Abel78} and Yao \cite{Yao79} proposed instead to measure complexity in terms of ``data movement'' between computing entities (processors) while ignoring local computations. In their interactive model, one entity knows $\mathbf{x}_{1}$ and another knows $\mathbf{x}_{2}$, both $\mathbf{x}_{1}$ and $\mathbf{x}_{2}$ being length $n$ (say, binary) vectors. The goal is for one of the entities to compute a given function $f(\mathbf{x}_{1},\mathbf{x}_{2})$. Complexity is then defined as the minimum number of exchanged bits between the two entities. Communication in this setup involves no coding in the sense that protocols between entities allow to compute the function after each instance of the sources---$\mathbf{x}_{1}$ and $\mathbf{x}_{2}$ represent one instance of source $1$ and one instance of source $2$, respectively. This framework lead the foundations of communication complexity and has been widely studied ever since (see, {\it{e.g.}}, \cite{PapaSip82}, \cite{BabaFra86,Orli90,Orli91,KushNis97}), though for ``simple'' networks involving no more than three nodes.

 Coding for computing over multiple source instances was first considered by Ahlswede and Cai \cite{AhlsCai94} for the Abelson-Yao's setup. The non-interactive (one-way) version was subsequently considered by Alon and Orlitsky \cite{AlonOrl96} and Koulgi et al. \cite{KoulTun03}. Recently, Shayevitz \cite{Shay11} investigated function computation over a cascade network where the transmitter can communicate to the receiver only via a relay.
 
\iffalse In the previous works, the network topology is fixed and analysis is carried out in the limit of large source instances. A different line of research was pursued by the Giridhar and Kumar \cite{GiriKum05} who provided scaling laws on the maximum network computation rate in terms of the number of nodes in the network.\footnote{Network computation rate refers to the number of computations per transmitted message in the network, typically assuming a common data rate across communication links.} Analysis was restricted to certain classes of functions and assumes a common data rate over each communication link.  
\fi

Close to our setting is the one of Kowshik and Kumar  \cite{KowsKum12} who investigated function computation over rooted directed trees and rooted directed acyclic graphs in which no interaction is allowed. Main results for rooted directed trees are necessary and sufficient conditions on the nodes' encoding procedures that allow function computation error free. When the sources' distribution is positive, these conditions are independent and allow to compute the rate region. For more general distributions these conditions appear hard to translate into bounds on the rate region.  

Another closely related work is the one of Appuswamy et al. \cite{AppuMasNik10,AppuFraNik13} who derived bounds on the maximum network computation rate for general directed acyclic graphs and independent sources.\footnote{An extension to multiple receivers was considered by Kannan and Viswanath \cite{KannVis13}.}

\subsubsection*{\sc Asymptotic zero-error probability}

Slepian and Wolf \cite{SlepWol73} characterized the rate region for multiple access networks and the identity function, \textit{i.e.}, when the receiver wants to recover the sources. For non-identity functions the problem was considered by K\"orner and Marton  \cite{KornMar79} who investigated the problem of computing the sum modulo two problem of two binary sources.  The rate region was established only for the case of symmetric distributions and was obtained by means of Elias's linear scheme \cite{elia55}. Variations of this scheme have later been used for computing linear functions over multiple access networks (see, {\it{e.g.}}, \cite{AhlsHan83}, \cite{HanKob87}, \cite{LaliPraVin11}, \cite{HuanSko12}).

An early and perhaps less known paper of Gel'fand and Pinsker \cite{GelfPin79} provides bounds for multiple access networks and arbitrary functions. They showed that these bounds are tight in a general case which includes the case of (conditionally) independent sources. As a byproduct they derived the optimal compression rate for the single source and arbitrary function setting with side information at the receiver. For this latter, an equivalent solution in terms of graph entropy was established by Orlitsky and Roche \cite{OrliRoc01}. This graph entropy approach was later used for multiple access networks in \cite{SefiTchISIT11} 
and in \cite{SefiTch12} for the case of cooperative transmitters.\footnote{An early work on multiple access with cooperative transmitters is \cite{EricKor83}.} 

In addition to multiple access networks, function computation over cascade networks have been investigated in \cite{CuffSuElg09}, \cite{Visw10}, and \cite{SefiTch122} referenced here in increasing order of generality. 

Beyond multiple access and cascade networks, collocated networks have been investigated by Ma, Ishwar and Gupta \cite{MaIshGup09} who established the rate region for independent sources.

Function computation over general networks remains challenging.  As summarized in \cite{KowsKum12} such problems ``\textit{combine the complexity of source coding of correlated sources with
rate distortion, together with the complications introduced by
the function structure.}'' Our results provide further insights by establishing the rate region for a general class of networks with possibly dependent sources.

\iffalse
The asymptotic zero error probability criterion is less stringent and may lead to larger rate regions by taking advantage of the correlation between sources. However, this usually comes at the expense of delay because of the use of long codewords. When delay is an issue, considering the zero-error criterion for finite block length may be more suitable.
\fi

The paper is organized as follows.  Section~\ref{sec:treePrelim}  provides graph related preliminaries and Section~\ref{sec:treeProbSetup} contains the precise problem formulation.  Results are presented in Section \ref{sec:treeResults} and their proofs are given in Section~\ref{sec:treeAnalysis}. 

\section{Preliminaries: tree, characteristic graph, and graph entropy} \label{sec:treePrelim}
We provide some graph theoretic background and introduce various notations which are summarized in Table~\ref{table} to come.
  
We denote by  $\mathcal{V}(G)$ and   $\mathit{E}(G)$ the vertex set and the edge set, respectively, of an undirected graph $G$. An undirected edge between nodes $u$ and $v$ is denoted by $\overline{uv}$ or  $\overline{vu}$. An \text{independent set} of a graph is a subset of its vertices no two of which are connected. A \text{maximal independent set} is an independent set that is not included in any other independent set.  The set of independent sets of a graph $G$ and the set of  maximal independent sets of $G$ are denoted by $\Gamma(G)$ and  $\Gamma^*(G)$, respectively.

A \text{path} between two nodes $u$ and $v$ in a given graph $G$ is a sequence of nodes  $u_1,\cdots,u_k$ where $u_1=u$, $u_k=v$, and $\overline{u_iu_{i+1}}\in \mathit{E}(G)$ for $1 \leq i \leq k-1$. A graph $G$ is \text{connected} if there exists a path between any two vertices $u,v \in \mathcal{V}(G)$. A path $u_1,\cdots,u_k$, $k\geq 2$, with $u_1=u_k$  is called a \text{cycle}. A graph is called \text{acyclic} if it contains no cycle. \iffalse nor self-loops, \textit{i.e.}, no edge of the form $\overline{uu}$. \fi A \text{tree} is a connected acyclic graph.

A \text{directed graph}, denoted by $\overrightarrow{G}$, is a graph whose edges have a direction. We use $\overrightarrow{uv}$ to denote an edge from node $u$ to node $v$. A \text{directed path} from node $u$ to node $v$ is a sequence of nodes  $u_1,\cdots,u_k$ where $u_1=u$, $u_k=v$ and $\overrightarrow{u_iu_{i+1}}\in \mathit{E}(\overrightarrow{G})$ for $1 \leq i \leq k-1$. 

The set of \text{incoming neighbors} of a node $u \in \mathcal{V}(\overrightarrow{G})$, denoted by $\mathit{In}(u)$, 
 is the set of nodes $v \in \mathcal{V}(\overrightarrow{G})$ such that $\overrightarrow{vu}\in \mathit{E}(\overrightarrow{G})$.  Their number, {\it{i.e.}}, $|\mathit{In}(u)|$, is sometimes variously denoted by $n(u)$. For a vertex $u \in 
\mathcal{V}(\overrightarrow{G})$, we denote  by $\mathit{Child}(u)$ the set of all nodes $v$ such that there exists a directed path from $v$ to $u$, including $u$ itself, and by $\mathit{Strangers}(u)$  the set of vertices $v$  for which there is no directed path between $u$ and $v$, \textit{i.e.},
$$\mathit{Strangers}(u) \defeq \{v: v \notin \mathit{Child}(u) \text{ and } u \notin \mathit{Child}(v)\}.$$ 
A \text{rooted directed tree},\footnote{Notice that in a rooted directed tree it is perhaps more common to consider edge directions from the root to the leaf by contrast with the present setup where information flows from the leaves to the root.} denoted by $\overrightarrow{T}$, is a directed tree where all the edges point towards the root node $r$.\footnote{Hence we have  $\mathit{Child}(r)=\mathcal{V}(\overrightarrow{T})$, {\it{i.e.}}, for any node $u \in \mathcal{V}(\overrightarrow{T})$ there exists a directed path from $u$ to $r$.} The immediate (unique) vertex which $u$ is pointing to is denoted by $u_{out}$, whenever $u\neq r$. 

 For a rooted directed tree $\overrightarrow{T}$, an \text{ordering} 
$$O_{\overrightarrow{T}}:\mathcal{V}(\overrightarrow{T}) \rightarrow \{1,2,\cdots,|\mathcal{V}(\overrightarrow{T})|\}$$ is a one-to-one mapping from the set of vertices to the natural numbers $\{1,2,\cdots,|\mathcal{V}(\overrightarrow{T})|\}$ such that if for two vertices $u,v \in \mathcal{V}(\overrightarrow{T})$ $$O_{\overrightarrow{T}}(u)>O_{\overrightarrow{T}}(v),$$ then the directed edge $\overrightarrow{uv}$ does not exist.\footnote{Note that an ordering imposes a (strict) total order on transmissions. Referring to Fig.~\ref{fig:rootedDirectedTree2}, information transmission occurs in three hops, first from nodes $\{1,2,3,4\}$, then nodes $\{5,6,7\}$, and finally nodes $\{8,9\}$. An ordering is obtained by first performing any three permutations separately on each of these sets, then concatenating the values of these sets, and finally adding the root node $10$. Valid orderings are thus, for example, $[O(i)]_{i=1,\ldots,10} =[1,2,3,4,5,6,7,5,9,8,10]$ (natural ordering), $[O(i)]_{i=1,\ldots,10} =[4,1,2,3,7,5,6,9,8,10]$, and $[O(i)]_{i=1,\ldots,10} =[2,1,3,4,6,7,5,8,9,10]$.}   The function  
$$O_{\overrightarrow{T}}^{-1}: \{1,\cdots,|\mathcal{V}(\overrightarrow{T})|\} \rightarrow \mathcal{V}(\overrightarrow{T})$$ 
denotes the inverse of  $O_{\overrightarrow{T}}$. 

For any vertex $u$ and any ordering $O_{\overrightarrow{T}}$,  $\mathit{Sub}_{O_{\overrightarrow{T}}}(u)$ and   $\mathit{Sup}_{O_{\overrightarrow{T}}}(u)$ denote the set of vertices with lower and higher orderings than $u$, respectively:
$$	\mathit{Sub}_{O_{\overrightarrow{T}}}(u)\defeq \{v:v\in \mathcal{V}(\overrightarrow{T}), O_{\overrightarrow{T}}(v) <O_{\overrightarrow{T}}(u)\}$$
$$	\mathit{Sup}_{O_{\overrightarrow{T}}}(u)\defeq \{v:v\in \mathcal{V}(\overrightarrow{T}), O_{\overrightarrow{T}}(v) >O_{\overrightarrow{T}}(u)\}.$$
In particular, we have
$$\{u\}\cup \mathit{Sub}_{O_{\overrightarrow{T}}}(u)\cup \mathit{Sup}_{O_{\overrightarrow{T}}}(u)=\mathcal{V}(\overrightarrow{T})$$
for any $u\in \mathcal{V}(\overrightarrow{T}) $.

Finally, for any vertex $u$ and any ordering $O_{\overrightarrow{T}}$ define 
\begin{align*}
\mathit{Roots}_{O_{\overrightarrow{T}}}(u)\defeq\{v: v \in \mathit{Sub}_{O_{\overrightarrow{T}}}(u),v_{out} \notin \mathit{Sub}_{O_{\overrightarrow{T}}}(u)\cup \{u\}\}
\end{align*}
{\it{i.e.}}, $\mathit{Roots}_{O_{\overrightarrow{T}}}(u)$ represents the set of nodes $v$ whose order is lower than $u$ but for which there exists no directed path from $v$ to $ \mathit{Sub}_{O_{\overrightarrow{T}}}(u)\cup \{u\}$. 

The definition of $\mathit{Roots}_{O_{\overrightarrow{T}}}(u)$ can be interpreted as follows. Consider the restriction of $\overrightarrow{T}$ to the set of vertices $ \mathit{Sub}_{O_{\overrightarrow{T}}}(u)\cup \{u\}$. This subgraph is composed of some disconnected rooted directed trees\footnote{A graph composed of a single node is considered a (degenerate) tree.} whose roots are $\mathit{Roots}_{O_{\overrightarrow{T}}}(u) \cup \{u\}$.

\begin{exam}
Consider the rooted directed tree $\overrightarrow{T}$ depicted in Fig.~\ref{fig:rootedDirectedTree2} with node  $r=10$ being the root. 
For vertex $2$, the unique outgoing neighbor is $5$ and the set of incoming neighbors is 
$$\mathit{In}(8)=\{5,6\}.$$
 Also, we have $\mathit{Child}(8)=\{1,2,5,6\}$ and $\mathit{Strangers}(8)=\{3,4,7,9\}$. 
 
 A possible ordering is the ordering given by the labels of the nodes (which already satisfies the ordering definition):
\begin{align*}
O_{\overrightarrow{T}}(i)=i\qquad 1 \leq i \leq 10.
\end{align*}
For this ordering we have 
\begin{align*}
\mathit{Sub}_{O_{\overrightarrow{T}}}(7)&=\{1,2,3,4,5,6\} \\
\mathit{Sup}_{O_{\overrightarrow{T}}}(7)&=\{8,9,10\} \\
\mathit{Roots}_{O_{\overrightarrow{T}}}(7)&=\{5,6\}.
\end{align*}
\end{exam}

Conditional characteristic graph plays a key role in coding for computing. We give here a general definition:
\begin{definition}[Conditional Characteristic Graph]\label{def:condCharGraph} Let $(L,K,S)\sim p(l,k,s)$ be a triplet of random variables taking on values over some finite alphabet $\mathcal{L}\times \mathcal{K}\times \mathcal{S}$. Let $f:\mathcal{S}\rightarrow \mathbb{R}$ be a function such that $H(f(S)|L,K)=0$. The conditional characteristic
graph $G_{L|K}(f)$ of $L$ given $K$ with respect to $f(s)$ is the graph whose vertex set is
$\mathcal{L} $ and such that $l_1 \in \mathcal{L}$ and $l_2 \in \mathcal{L}$ are connected if for some $s_1,s_2 \in \mathcal{S}$,  and $k \in \mathcal{K}$
 \begin{itemize} 
\item[i.] $p(l_1,k,s_1) \cdot p(l_2,k,s_2) > 0$, 
\item[ii.] $f(s_1)\neq f(s_2)$. 
\end{itemize}
When $f(s)$ is known by the context, the above conditional characteristic graph is simply denoted by $G_{L|K}$.
\end{definition}
\begin{remark}
 When $L=S=X$ and $K=\varnothing$, Definition~\ref{def:condCharGraph} reduces to the definition of the characteristic graph introduced by K\"orner in \cite{Korn73} and when $S=(X,Y)$, $L=X$, and $K=Y$ Definition~\ref{def:condCharGraph} reduces to the definition of conditional characteristic graph introduced by Witsenhausen in \cite{Wits76}. 
\end{remark}
Definition \ref{def:condCharGraph} can be interpreted as follows. Suppose a transmitter has access to random variable $L$ and a receiver has access to random variable $K$ and wants to compute function $f(S)$. The condition $H(f(S)|L,K)=0$ guarantees that by knowing $L$ and $K$ the receiver can compute $f(S)$. Moreover, in the characteristic graph $G_{L|K}$, given $K=k,$ the knowledge of an independent set of $G_{L|K}$ that includes the
realization $L=l$ suffices for the receiver to compute $f(S)$ since no two vertices in an independent set can produce different function outputs. Hence, for computing $f(S)$ the receiver needs only to know an independent set that includes $L$.

\begin{exam}  \label{ex:bigger1}
Let $ X $ and $ Y $  be random 
variables defined over the alphabets $ \mathcal{X}$ and 
$\mathcal{Y}$, respectively, with  $$ \mathcal{X}=\mathcal{Y}=\{1,2,3,4\}.$$
Further, suppose that  $ P(X=Y)=0 $ and that $(X,Y)$ takes on values uniformly over the pairs $ (i,j) \in \mathcal{X} \times 
\mathcal{Y}$ with $ i \neq j $. Let $f(x,y)$ be defined as
 \begin{gather}
\raisetag{-10pt}
f(x,y) = \begin{cases}
0 \mbox{  if } x<y,\\
1  \mbox{ if } x>y. 
\end{cases}  \nonumber
\end{gather}

\begin{figure}
\begin{center}
\begin{tikzpicture}[scale=0.9]
\draw [black,fill=black] (0,0) circle [radius=0.05];
\draw [black,fill=black] (2,0) circle [radius=0.05];
\draw [black,fill=black] (0,2) circle [radius=0.05];
\draw [black,fill=black] (2,2) circle [radius=0.05];
\node [above left] at (0,2) {$1$};
\node [above right] at (2,2) {$2$};
\node [below right] at (2,0) {$3$};
\node [below left] at (0,0) {$4$};

\draw [thick](0,0)--(2,2);
\draw [thick](0,0)--(0,2);
\draw [thick](0,2)--(2,0);

\end{tikzpicture}
\caption{$G_{X|Y}$.}
\label{fig:condGraph}
\end{center}
\end{figure}

In Definition~\ref{def:condCharGraph}, let $S=(X,Y)$, $L=X$, and $K=Y$. Fig.~\ref{fig:condGraph} depicts $ G_{X|Y}$ and we have
\begin{align*}
 \Gamma(G_{X|Y})=\{ \{1\},\{2\},\{3\},\{4\},\{1,2\},\{2,3\},\{3,4\} \}
\end{align*}
and
$$ \Gamma^*(G_{X|Y})=\{ \{1,2\},\{2,3\},\{3,4\} \}. $$
\end{exam}

In this example, the maximal independent sets overlap with each other and do not partition the vertices of the graph. The following lemma, whose proof is deferred to Appendix~\ref{app:condGraphIndependent}, provides a sufficient condition under which the set of maximal independent sets forms a partition of the vertices of $G_{L|K}$. 
\begin{lemma}\label{lem:condGraphIndependent}  Let 
$$(L,K,S_1,S_2)\sim p(l,k,s_1,s_2)=p(l,s_1)\cdot p(k,s_2)$$ and $f:\mathcal{S}_1 \times \mathcal{S}_2\rightarrow \mathbb{R}$ be a function such that $H(f(S_1,S_2)|L,K)=0$. Then $\Gamma^*(G_{L|K})$ is a partition of the set $\mathcal{L}$. In other words, each $l \in \mathcal{L}$ is included in exactly one maximal independent set.
\end{lemma}
 A multiset
of a set $\cal{S}$ is a collection of elements from ${\cal{S}}$ possibly with
repetitions, {\it{e.g.}}, if ${\cal{S}}=\{0,1\}$, then $\{0,1,1\}$ is a multiset. We use ${\text{M}}({\cal{S}})$ to denote the
collection of all multisets of $\cal{S}$.

\begin{definition} [Conditional Graph Entropy \cite{OrliRoc01}] \label{def:condGraphEntropy} Given $(L,K,S)\sim p(l,k,s)$ and $f:\mathcal{S}\rightarrow \mathbb{R}$  such that $\mathcal{S}$ is a finite set and $H(f(S)|L,K)=0$, the conditional graph entropy $H(G_{L|K}(f))$ is defined as\footnote{Given two random variables $X$ and $V$, where $X$ ranges over
$\cal{X}$ and $V$ over {\emph{subsets}} of $\cal{X}$ (\it{i.e.}, a sample
of $V$ is a subset of $\cal{X}$), we write $ X \in V $
whenever $P(X\in V)=1$.}
 \begin{align*}
H(G_{L|K}(f))\defeq \min \limits_{\substack{ V -
L- K\\ L \in V \in {\text{M}}(\Gamma(G_{L|K}(f)))}} I(V;L|K)=\min \limits_{\substack{ V -
L- K\\ L \in V \in \Gamma^*(G_{L|K}(f))}} I(V;L|K).
\end{align*}
\end{definition}

When the function $f(s)$ is known by the context, the above conditional graph entropy is simply denoted by  $H(G_{L|K})$. Note that we always have  $H(G_{L|K}(f)) \leq H(L|K)$.

\begin{exam}  \label{ex:bigger2}
Consider Example~\ref{ex:bigger1}. According to Definition~\ref{def:condGraphEntropy}, for computing $H(G_{X|Y})$ we can restrict the minimization of $I(V;X|Y)$ to be over all $V$ that take values over maximal independent sets, \textit{i.e.},
$$\mathcal{V}=\{v_1=\{1,2\},v_2=\{2,3\},v_3=\{3,4\}\}.$$
Moreover, from the condition $X \in V$ and the symmetries of the pair $(1,4)$ and the pair $(2,3)$ it can be deduced that the  
 $p(v|x)$ that minimizes the mutual information $I(V;X|Y)$ is given by
\begin{align*}
p(v_1|2)&=p(v_3|4)=\delta \\
p(v_2|2)&=p(v_2|3)=1-\delta,\\
p(v_1|1)&=p(v_3|4)=1 \\
p(v_2|1)=p(v_3|1)&=p(v_1|4)=p(v_2|4)=0 
\end{align*}
for some $\delta \in [0,1]$. This gives
$$I(V;X|Y)=\frac{1}{2}\cdot (H(\delta/3,(1+\delta)/3,(2-2\delta)/3)+H(1/3,(1-\delta)/3,(1+\delta)/3)-h_b (\delta))$$ 
where $h_b(\delta)$ denotes the binary entropy $-\delta \log \delta-(1-\delta)\log (1-\delta)$.
It can be checked that $I(V;X|Y)$ is minimized for $\delta=1$. Hence, $H(G_{X|Y})\simeq 0.92 <1.58 \simeq H(X|Y)$ and the alphabet of the optimal $V$ is $\mathcal{V}=\{v_1=\{1,2\},v_3=\{3,4\}\}$ since $p(v_2|2)=p(v_2|3)=1-\delta=0$.
\end{exam}

The following table summarizes the main notations used throughout the paper.

{\begin{center}
\begin{tabular}{|c|c|}
\hline
Notation & Definition \\ 
\hline
$G$ &  Graph  \\ 
\hline
$\mathcal{V}(G)$ & Set of vertices of $G$  \\ 
\hline
$\mathit{E}(G)$ & Set of edges of $G$  \\ 
\hline 
$\Gamma(G)$ & Set of independent sets of $G$  \\ 
\hline 
$\Gamma^* (G)$ & Set of maximal independent sets of $G$ \\ 
\hline 
$\text{M}(\Gamma (G))$ & Multiset of independent sets of $G$ \\ 
\hline 
$\overrightarrow{T}$ & A rooted directed tree  \\ 
\hline
$r$ & Root of a rooted directed tree  \\ 
\hline
$\mathit{In}(u)$  & Set of vertices whose outgoing edges directly point to vertex $u$ \\ 
\hline
$u_{out}$ &  The outgoing neighbor of vertex $u$ \\ 
\hline 
$\mathit{Child}(u)$ & Set of vertices with directed path to $u$, including $u$ itself\\
\hline
$\mathit{Strangers}(u)$  & Set of vertices $v$ with no directed path between $u$ and $v$\\
\hline 
 $O_{\overrightarrow{T}}$ & An ordering\\ 
\hline
$\mathit{Sub}_{O_{\overrightarrow{T}}}(u)$ & Set of vertices with lower ordering than $u$\\ 
\hline 
  $\mathit{Sup}_{O_{\overrightarrow{T}}}(u)$ & Set of vertices with higher ordering than $u$ \\ 
\hline
  $\mathit{Roots}_{O_{\overrightarrow{T}}}(u)$ & Roots (except from $u$) of the restriction of $\overrightarrow{T}$ to the set of vertices $ \mathit{Sub}_{O_{\overrightarrow{T}}}(u)\cup \{u\}$ \\ 
\hline
$G_{L|K}(f)$  & Conditional characteristic graph of $L$ given $K$ with respect to function $f$  \\ 
\hline
   $H(G_{L|K}(f))$  & Conditional graph entropy \\ 
\hline
\end{tabular} 
\end{center}
\captionof{table}{Notation list \label{table}}
}
\section{Problem formulation}\label{sec:treeProbSetup}
Consider a rooted directed tree $\overrightarrow{T}$ with root $r$. Let\footnote{In general, for a set $\mathcal{A}$, we define $\mathcal{X}_{\mathcal{A}}\defeq (\mathcal{X}_u: u \in \mathcal{A})$.} $$\mathcal{X}_{\mathcal{V}(\overrightarrow{T})}\defeq (\mathcal{X}_u: u \in \mathcal{V}(\overrightarrow{T}))$$ and $$f:\mathcal{X}_{\mathcal{V}(\overrightarrow{T})}\rightarrow \mathcal{F}$$ where $\mathcal{X}_{u}, u \in \mathcal{V}(\overrightarrow{T})$ are finite sets. Node $u \in \mathcal{V}(\overrightarrow{T})$ has access to random variable $X_u\in {\mathcal{X}}_u$.  Let $\{(x_{\mathcal{V}(\overrightarrow{T})})_i \}_{i\geq 1}$,  be independent instances of
random variables $X_{\mathcal{V}(\overrightarrow{T})}$ taking values
over $ \mathcal{X}_{\mathcal{V}(\overrightarrow{T})}$ and
distributed according to $p(x_{\mathcal{V}(\overrightarrow{T})})$.

To simplify notation, in the following we shall often avoid any explicit reference to the underlying tree $\overrightarrow{T}$ and will write, for instance, simply $O$ and  $\mathcal{V}$ instead of $O_{\overrightarrow{T}}$ and $\mathcal{V}(\overrightarrow{T})$, respectively.

\begin{definition}[Code]
A $((2^{nR_{u}})_{u \in \mathcal{V}\setminus\{ r\}},n)$ code consists of encoding functions \begin{align*} \varphi_{u}:\mathcal{X}_u^n  \otimes_{v \in \mathit{In}(u)}  \{ 1,\cdots,2^{nR_{v}} \} \rightarrow  \{ 1,\cdots,2^{nR_{u}} \} 
\end{align*} 
at nodes $u\in \mathcal{V}\setminus \{r\}$ and a decoding function
\begin{align*}  
\psi&:\mathcal{X}_r^n \otimes_{v \in \mathit{In}(r)} \{1, \cdots, 2^{nR_{v}}\} \rightarrow \mathcal{F}^n \nonumber 
\end{align*} 
at the root $r$.
 
Recall that by definition of  $\mathit{Child}(u)$ we have  $$\mathit{Child}(u)=\{u\}\bigcup \limits_{v \in \mathit{In}(u)} \mathit{Child}(v).$$This allows to recursively define 
$$\varphi_{u}(\mathbf{X}_{\mathit{Child}(u)}) \defeq \varphi_{u}(\mathbf{X}_{u},\varphi_{u_{1}}(\mathbf{X}_{\mathit{Child}(u_{1})}),\cdots, \varphi_{u_{n(u)}}(\mathbf{X}_{\mathit{Child}(u_{n(u)})})),$$
where $\{u_{1},u_{2},\cdots,u_{n(u)}\}=\mathit{In}(u)$.

Throughout the paper we use bold fonts to denote length $n$ vectors. In the above expression, for instance, $\mathbf{X}_{\mathit{Child}(u)}$ denotes a block of $n$ independent realizations of ${X}_{\mathit{Child}(u)}$.  

 The (block) error probability of a code (averaged over the sources' outcomes) is defined as
$$P(\psi(\mathbf{X}_r,\varphi_{r_{1}}(\mathbf{X}_{\mathit{Child}(r_{1})}),\cdots, \varphi_{r_{n(r)}}(\mathbf{X}_{\mathit{Child}(r_{n(r)})}))\ne f(\mathbf{X}_{\mathcal{V}}))$$
where $\{r_{1},r_{2},\cdots,r_{n(r)}\}=\mathit{In}(r)$ and where with a slight abuse of notation we wrote $f(\mathbf{X}_{\mathcal{V}})$ to denote $n$ (independent) realizations of $f(X_{\mathcal{V}}$).
\end{definition}
\begin{definition}[rate region]
A rate tuple $(R_{u})_{u \in \mathcal{V}\setminus\{ r\}}$ is achievable if, for any $\varepsilon > 0$ and all $n$
large enough, there exists a $((2^{nR_{u}})_{u \in \mathcal{V}\setminus\{ r\}},n)$ code whose error probability is no
larger than $\varepsilon$. The rate region is the closure of the set of achievable
rate tuples $(R_{u})_{u \in \mathcal{V}\setminus\{ r\}}$.
\end{definition}
In this paper we seek to characterize the rate
region for given $\overrightarrow{T}$, $f$, and~$p(x_{\mathcal{V}})$.
 
\section{Results} \label{sec:treeResults}
We start with a cut-set outer bound to the rate region. Here, a valid cut is a subset $\mathcal{S} \subset \mathcal{V}$ such that if $u \in \mathcal{S}$ then $\mathit{Child}(u)\subseteq \mathcal{S}$.  
\begin{theorem}[Outer Bound] \label{th:treeUpper} 
If a rate tuple $(R_{u})_{u \in \mathcal{V}\setminus\{ r\}}$ is achievable, then for any valid cut $\mathcal{S} \subset \mathcal{V}$, we have
\begin{align}
&\sum \limits_{\substack{v \in \mathcal{S}:\\v_{out} \in \mathcal{S}^c}}R_{v} \geq  H(G_{X_{\mathcal{S}}|X_{\mathcal{S}^c}}). \nonumber
\end{align} 
\end{theorem}
The above result is an immediate extension of the single source result  \cite[Theorem~1]{OrliRoc01}. It can also be easily checked that the above outer bound implies \cite[Corollary~2]{SefiTchISIT11}  when $\overrightarrow{T}$ is a multiple access network and implies \cite[Theorem 2]{Visw10}  when $\overrightarrow{T}$ is a cascade network.

\begin{figure}
\begin{center}
\begin{tikzpicture}[scale=0.8]

\draw [black,fill=black] (4,0) circle [radius=0.05];
\draw [black,fill=black] (6,0) circle [radius=0.05];
\draw [black,fill=black] (3,2) circle [radius=0.05];
\draw [black,fill=black] (5,2) circle [radius=0.05];
\draw [black,fill=black] (4,4) circle [radius=0.05];
\draw [black,fill=black] (3,6) circle [radius=0.05];
\draw [black,fill=black] (1,2) circle [radius=0.05];

\draw [->][thick](4.1,0.2)--(4.9,1.8);
\draw [->][thick](3.9,4.2)--(3.1,5.8);
\draw [->][thick](4.9,2.2)--(4.1,3.8);
\draw [->][thick](5.9,0.2)--(5.1,1.8);

\node [below left] at (4,0) {$\mathbf{X}_3$};
\node [below right] at (6,0) {$\mathbf{X}_{4}$};
\node [left] at (1,2) {$(\mathbf{X}_5,\mathbf{W}_1,\mathbf{W}_2)$};
\node [right] at (3,2) {$\mathbf{X}_6$};
\node [right] at (5,2) {$\mathbf{X}_7$};
\node [right] at (4,4) {$\mathbf{X}_9$};
\node [above] at (3,6) {$\mathbf{X}_{10}$};
\node [left] at (2,4) {$\mathbf{X}_8$};

\node [left] at (4.5,1) {\small{$R_3$}};
\node [right] at (5.5,1) {\small{$R_{4}$}};
\node [left] at (1.5,3) {\small{$R_5$}};
\node [right] at (2.5,3) {\small{$R_6$}};
\node [right] at (4.5,3) {\small{$R_7$}};
\node [left] at (2.5,5) {\small{$R_8$}};
\node [right] at (3.5,5) {\small{$R_9$}};

\draw [->][thick](1.1,2.2)--(1.9,3.8);
\draw [->][thick](2.9,2.2)--(2.1,3.8);
\draw [black,fill=black] (2,4) circle [radius=0.05];
\draw [->][thick](2.1,4.2)--(2.9,5.8);

\end{tikzpicture}
\end{center}
\caption{Resulting tree after the first iteration of the achievable scheme applied on the tree depicted in Fig.~\ref{fig:rootedDirectedTree2}.}
\label{fig:rootedDirectedTreeStep2}
\end{figure}
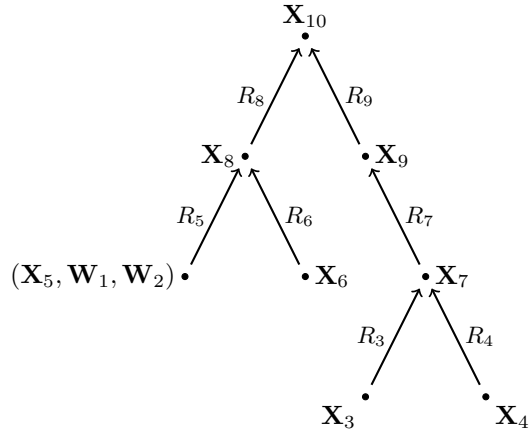
Theorem~\ref{th:treeAchieve} to come provides an inner bound to the rate region. For a given ordering, the scheme used for establishing this inner bound applies the scheme proposed in \cite[Proof of Proposition~1]{SefiTchISIT11} for the multiple access configuration in an iterative fashion.  To describe the main idea, consider the network depicted in Fig.\ref{fig:rootedDirectedTree2} where $f$ is a function of $X_1^{10}$ and consider the natural (valid) ordering given by the labels of the nodes. \iffalse Using our convention, we denote by $\bm{f}({X}_1^{10})$ a block of $n$ realizations of $f(X_1^{10})$. \fi\begin{description}
\item[Step $1$:] Vertex $1$ chooses a message $\mathbf{W}_1\in {\cal{W}}_1^n$ such that each realization of ${f}({X}_1^{10})$ is computable from the corresponding values in $\mathbf{W}_1$ and $\mathbf{X}_2^{10}$.\footnote{Note that one alternative choice for $\mathbf{W}_1$ would be to have $\mathbf{W}_1\mathbf{X}_1$. However, this may not be efficient. In the proposed scheme $\mathbf{W}_1$ is chosen as a block of independent sets of some proper characteristic graph.} Vertex $2$  chooses a message $\mathbf{W}_2\in {\cal{W}}_2^n$ such that each realization of ${f}({X}_1^{10})$ is computable from the corresponding values of $\mathbf{W}_1$, $\mathbf{W}_2$, and $\mathbf{X}_3^{10}$. 

\item[Step $2$:]  Both vertices $1$ and $2$ transmit their messages to vertex $5$ through a Slepian-Wolf coding that allows vertex $5$ to decode $\mathbf{W}_1$ and $\mathbf{W}_2$ by having access to the side information $\mathbf{X}_5$.  

\item[Step $3$:] Remove vertices $1$ and $2$ and all edges connected to them, and replace $\mathbf{X}_5$ by $(\mathbf{X}_5,\mathbf{W}_1,\mathbf{W}_2)$. The resulting tree is depicted in Fig. \ref{fig:rootedDirectedTreeStep2}. 
\item[Step $4$:] Repeat Steps $1$, $2$, and $3$ until the root receives the messages $\mathbf{W}_8$ and $\mathbf{W}_9$ from which it can compute the function reliably.
\end{description}
\begin{theorem}[Inner Bound] \label{th:treeAchieve}
An inner bound to the rate region is the convex hull of the rate tuples $(R_{u})_{u \in \mathcal{V}\setminus\{ r\}}$ such that 
\begin{align}
&\sum \limits_{v \in \mathcal{S}}R_{v} \geq I(X_{\mathcal{S}},W_{\mathit{In}(\mathcal{S})};W_{\mathcal{S}}|X_{u},W_{\mathcal{S}'})\label{eq:treeAchieveRates} \\   u \in   \mathcal{V},\quad \varnothing  \neq & \mathcal{S} \subseteq \mathit{In}(u), \quad
\mathcal{S}'=\mathit{In}(u) \setminus \mathcal{S}, \quad \mathit{In}(\mathcal{S})\defeq\bigcup \limits_{v \in \mathcal{S}}\mathit{In}(v), \nonumber
\end{align} 
where random variables $(W_u)_{u\in \mathcal{V}\setminus \{r\}}$  satisfy
the Markov chain conditions 
\begin{align}
W_u-(X_{u},W_{\mathit{In}(u)})-(X_{\mathit{Child}(u)^c},W_{\mathit{Strangers}(u)}), \label{eq:treeAchieveMarkovs2}
\end{align}
as well as the condition
\begin{small}
\begin{align}
(X_{u},W_{\mathit{In}(u)}) \in W_u \in \textnormal{M}(\Gamma(G_{X_{u},W_{\mathit{In}(u)}|X_{\mathit{Sup}_O(u)},W_{\mathit{Roots}_O(u)}})), \label{eq:treeAchieveSets}
\end{align}
\end{small}
 for an ordering $O$.  Moreover, the inner bound is the same regardless of the ordering O.
 \end{theorem}
Note that in the above iterative strategy, transmissions at any given node depend on the ordering. For instance, another possible ordering is the one obtained by swapping nodes $1$ and $2$ in Fig.\ref{fig:rootedDirectedTree2}, {\it{i.e.}}, $O(1)=2$, $O(2)=1$, and $O(i)=i$, $i\in \{3,4,\ldots,10\}$. For this ordering, Vertex $2$ chooses a message $\mathbf{W}_2$ such that each realization of ${f}({X}_1^{10})$ is computable from the corresponding values of $\mathbf{W}_2$ and $(\mathbf{X}_1,\mathbf{X}_3^{10})$. As a consequence, it may seem that the rate region achieved by the strategy depends on the ordering we impose on transmissions. As claimed in Theorem~\ref{th:treeAchieve}, the rate region is the same regardless of the ordering.
Indeed, later we shall see that if a set of auxiliary random variables satisfies \eqref{eq:treeAchieveMarkovs2} and \eqref{eq:treeAchieveSets} for a specific ordering, then it also satisfies these equations for any other ordering. Since \eqref{eq:treeAchieveRates} is independent of the ordering, this means that any two orderings give the same achievable rate tuples. 

Let us explain the terms \eqref{eq:treeAchieveRates}, \eqref{eq:treeAchieveMarkovs2}, and \eqref{eq:treeAchieveSets}. Random variable ${W}_u$ is interpreted as the message sent by vertex $u$ and the Markov condition~\eqref{eq:treeAchieveMarkovs2} reflects the fact that this message can depend only on the available side information $X_u$ and the set of incoming messages $W_{\mathit{In}(u)}$. Once vertex $u$ has transmitted its data, the aggregate information in the resulting tree  is $({W}_u,{X}_{\mathit{Sup}_O(u)},{W}_{\mathit{Roots}_O(u)})$. Choosing the alphabet of the message ${W}_u$ as in \eqref{eq:treeAchieveSets} guarantees that the knowledge of ${W}_u$ and 
 $({X}_{\mathit{Sup}_O(u)},{W}_{\mathit{Roots}_O(u)})$ suffices for computing $f$ error free. Finally, the rate condition \eqref{eq:treeAchieveRates} guarantees that  ${W}_u$ can be reliably  decoded at the outgoing neighbor of vertex $u$.

Note that in the above theorem, $W_u$ are not restricted to take values over maximal
independent sets. By contrast with the single transmitter case where the restriction to maximal
independent sets induces no loss of optimality (see \cite{OrliRoc01} and Definition~\ref{def:condGraphEntropy} where $V$ may be restricted to range over $\Gamma^*(G_{L|K}(f))$), for more than one transmitter the restriction to maximal independent sets may
indeed induce a loss of optimality. This was shown in an example in \cite{SefiTchISIT11} related to the multiple access network configuration.

Theorem~\ref{th:treeAchieve} recovers the inner bounds \cite[Proposition~1]{SefiTchISIT11} for the multiple access and \cite[Theorem~4]{SefiTch122} for the cascade network. 

\iffalse[ However, this region does not include \cite[Theorem~5]{SefiTch122} and consequently, \cite[Theorem~6]{SefiTch122} (which is a combination of \cite[Theorem~4]{SefiTch122} and \cite[Theorem~5]{SefiTch122}) for cascade network. The corresponding schemes for these inner bounds are proposed for cascade network which has only three nodes. The extension of these schemes to an arbitrary rooted directed tree seems to be difficult.]
\fi
\begin{remark} \label{remark:ProofOfClaimMarkovStrange}
The Markov chains \eqref{eq:treeAchieveMarkovs2} are equivalent to the following Markov chains
\begin{align}
W_u-(X_{u},W_{\mathit{In}(u)})-(X_{\mathit{Child}(u)^c},W_{\mathit{Sub}_O(u) \setminus \mathit{Child}(u)})\qquad u\in \mathcal{V}\setminus \{r\}, \label{eq:treeAchieveMarkovs}
\end{align}
for any ordering $O$. This equivalence shall prove useful for establishing Theorem~\ref{th:treeRate} to come. The proof of this remark is deferred to Appendix~\ref{subSec:ProofOfClaimMarkovStrange}. 
\end{remark}

The main result, stated in Theorem~\ref{th:treeRate} to come, characterizes the rate region when the sources satisfy the following Markov property:
\begin{definition}[Markov Property] \label{def:MarkovProperty}
Consider a vertex  $u$ in a rooted directed tree with sources $X_{\mathcal{V}}$ available at its nodes. Remove vertex $u$ from the tree together with its incoming and outgoing edges. The resulting graph is locally Markovian if the remaining sets of connected sources are independent given the value of $X_u$, {\it{i.e.}}, if
\begin{align*}
(X_{\mathit{Child}(u_1)},\cdots,X_{\mathit{Child}(u_{n(u)})},X_{\mathit{Child}(u)^c})
\end{align*}
are independent given $X_u$,  where  $\{u_1,\cdots, u_{n(u)}\}=\mathit{In}(u)$. 

A directed tree satisfies the Markov Property if it is locally Markovian for every $u \in \mathcal{V}$.\end{definition}

\begin{remark} It can be verified that a rooted directed tree satisfies the Markov property if and only if the joint probability distribution of  $X_{\mathcal{V}}$ is of the form
\begin{align}p(x_{\mathcal{V}})=p(x_r)\cdot \prod \limits_{u \neq r}p(x_u|x_{u_{out}}).
\end{align}\label{distprod}
The Markov Property thus holds, in particular, when all the sources $X_{\mathcal{V}}$ are independent.
\end{remark}

 \begin{theorem} \label{th:treeRate} For a rooted directed tree $\overrightarrow{T}$ that satisfies the Markov property, the inner and outer bounds given by Theorem~\ref{th:treeUpper} and Theorem~\ref{th:treeAchieve} are tight and the rate region is the set of all rate tuples $(R_{u})_{u \in \mathcal{V}\setminus\{ r\}}$ such that
\begin{align}
R_{u} \geq H(G_{X_{{\mathit{Child}(u)}}|X_{\mathit{Child}(u)^c}}) \qquad u \in  \mathcal{V}\setminus \{r\}.
\end{align}
 \end{theorem} 
 
As alluded to in the introduction, Theorem~\ref{th:treeRate} recovers all previously known rate regions related to network configurations with no interaction:
\begin{itemize}
\item for point-to-point we recover \cite[Theorem~2]{GelfPin79} and \cite[Theorem~1]{OrliRoc01} which characterize the rate region for arbitrary function and sources' probability distribution;
\item for the multiple access network we recover \cite[Theorem~2]{GelfPin79} which characterizes the rate region for arbitrary functions provided the sources at the transmitters are independent conditioned on the source at the receiver. We note in passing that \cite[Theorem~2]{GelfPin79} is stated with respect to auxiliary random variables whose range is left unspecified. By contrast, our rate region characterization is in terms of explicit auxiliary random variables---they take values over independent sets of some suitable characteristic graphs;
\item for the cascade network we recover \cite[Theorem~3]{Visw11} and \cite[Theorem~2]{SefiTch122} which derive the rate region for arbitrary functions when the sources form a Markov chain. Note that this result includes the result of \cite[Section~V.B.]{CuffSuElg09} which holds for the case where no side information is available at the receiver.
\end{itemize}

The following corollary essentially follows from Lemma~\ref{lem:condGraphIndependent} and Theorem~\ref{th:treeRate}:
 \begin{corollary}\label{th:cor}
 For a rooted directed tree $\overrightarrow{T}$ with independent sources $X_{\mathcal{V}}$,  the rate region is given by
  $$ R_{u} \geq H(W_u^*) \hspace{1 cm} u \in  \mathcal{V}\setminus \{r\},$$
where $\mathcal{W}^*_u\defeq \Gamma^*(G_{X_{\mathit{Child}(u)}|X_{\mathit{Child}(u)^c}})$ and for any $w^*_u \in \mathcal{W}^*_u$ $$p(W_u^*=w^*_u)\defeq \sum \limits_{ x_{\mathit{Child}(u)}\in w^*_u }p(x_{\mathit{Child}(u)}).$$
 \end{corollary}

\section{Proofs} \label{sec:treeAnalysis}

Throughout the section we often make use of robust typicality instead of the perhaps more standard use of weak/strong typicality. Robust typicality and its properties are recalled in Section~\ref{sect:typ} of the Appendix.
 
For notational simplicity we shall leave out any explicit reference to the ordering and write, for instance, $\mathit{Sub}(u)$ instead of $\mathit{Sub}_O(u)$. The order shall be understood from the context.

\begin{IEEEproof}[Proof of Theorem~\ref{th:treeUpper}]
Reveal $X_{\mathcal{S}}$ to all vertices in $\mathcal{S}$ and reveal $X_{\mathcal{S}^c}$ to all vertices in $\mathcal{S}^c$. Since each vertex in $\mathcal{S}$ has access to the same information, and since this is also holds for the vertices in $\mathcal{S}^c$, the sum rate constraint for the links from $\mathcal{S}$ to $\mathcal{S}^c$ is greater than or equal to the rate constraint where only one of the vertices in $\mathcal{S}$ communicates to one of the vertices in $\mathcal{S}^c$. Using the single source result \cite[Theorem~1]{OrliRoc01} completes the proof. 
\end{IEEEproof}

\begin{IEEEproof}[Proof of Theorem~\ref{th:treeAchieve}] 
Suppose random variables $W_{\mathcal{V}\setminus \{r\}}$ satisfy \eqref{eq:treeAchieveMarkovs2} and \eqref{eq:treeAchieveSets}. These random variables together with $X_{\mathcal{V}}$ are distributed according to some $p(x_{\mathcal{V}},w_{\mathcal{V}\setminus \{r\}})$.

For each $u \in \mathcal{V}\setminus \{r\}$, independently generate $2^{nR_{W_u}}$ sequences 
$$\mathbf{w}_{u}^{(i)}=(w_{u,1}^{(i)},w_{u,2}^{(i)},\ldots,w_{u,n}^{(i)}) \hspace{.5 cm}
i\in \{1,2,\ldots,2^{nR_{W_u}}\},$$ 
in an i.i.d. manner according to the marginal 
distribution $p(w_u)$; randomly and uniformly bin these
sequences into $2^{nR_{u}}$ bins; and reveal the bin assignments $\phi_{u}$ to vertices $u$ and $u_{out}$.
 
\noindent\textit{Encoding/decoding at intermediate nodes and leaves:} Given an ordering $O$, the encoding is done sequentially at vertices 
$$O^{-1}(1), O^{-1}(2), \cdots, O^{-1}(|\mathcal{V}|-1).$$ 
Let 
$$0<\varepsilon_1 < \varepsilon_2 < \cdots<\varepsilon_{|\mathcal{V}|-1}<\varepsilon_{|\mathcal{V}|}=\varepsilon\leq 1.$$

We distinguish leaves from intermediate nodes.\footnote{By intermediate node we intend any node that is not the root or a leaf.} 

If $u$ is a leaf, {\it{i.e.}}, $\mathit{In}(u)=\varnothing$, the corresponding encoder finds a sequence $\mathbf{w}_u$ such that  
 $$(\mathbf{x}_u,\mathbf{w}_u)\in  \mathcal{A}_{\varepsilon_i}^{(n)}(X_u,W_u)\quad \text{where} \quad i=O(u) $$ 
 and sends the index of the bin that contains it, {\it{i.e.}},
$\phi_{u}(\mathbf{w}_u)$, to  vertex $u_{out}$.

If $u$ is not a leaf, then the corresponding decoder first  decodes the set of $n(u)$ incoming messages as follows. Given  $\mathbf{x}_u $ and the incoming messages' indices 
$$(\phi_{u_{1}}(\mathbf{w}_{u_{1}}),\cdots,\phi_{u_{n(u)}}(\mathbf{w}_{u_{n(u)}})),$$ 
where $\{u_1,u_2,\cdots,u_{n(u)}\}=\mathit{In}(u)$, vertex $u$ declares
$$(\hat{\mathbf{w}}_{\mathit{In}(u)})$$
if it is the unique $(\hat{\mathbf{w}}_{\mathit{In}(u)})$ such that 
$$(\mathbf{x}_u,\hat{\mathbf{w}}_{\mathit{In}(u)}) \in 
\mathcal{A}_{\varepsilon_i}^{(n)}(X_u,W_{\mathit{In}(u)})\quad \text{where}\quad\quad i=O(u)$$ and such that
$$(\phi_{u_{1}}(\hat{\mathbf{w}}_{u_{1}}),\cdots,\phi_{u_{n(u)}}(\hat{\mathbf{w}}_{u_{n(u)}}))=(\phi_{u_{1}}(\mathbf{w}_{u_{1}}),\cdots,\phi_{u_{n(u)}}(\mathbf{w}_{u_{n(u)}})).$$
Having decoded $\hat{\mathbf{w}}_{\mathit{In}(u)}$, vertex $u$ finds a sequence $\mathbf{w}_u$ such that $$(\mathbf{x}_u,\hat{\mathbf{w}}_{\mathit{In}(u)},\mathbf{w}_u) \in 
\mathcal{A}_{\varepsilon_i}^{(n)}(X_u,W_{\mathit{In}(u)},W_u)$$  and sends the index of the bin that contains it, {\it{i.e.}},
$\phi_{u}(\mathbf{w}_u)$, to vertex $u_{out}$.

\noindent\textit{Decoding at the root:} Given  $\mathbf{x}_r $ and the incoming messages' indices 
$$(\phi_{r_{1}}(\mathbf{w}_{r_{1}}),\cdots,\phi_{r_{n(r)}}(\mathbf{w}_{r_{n(r)}})),$$ where  $\{r_{1},r_2,\cdots,r_{n(r)}\}=\mathit{In}(r)$, the root first declares
$$(\hat{\mathbf{w}}_{\mathit{In}(r)})$$
if it is the unique $(\hat{\mathbf{w}}_{\mathit{In}(r)})$ such that 
$$(\mathbf{x}_r,\hat{\mathbf{w}}_{\mathit{In}(r)}) \in 
\mathcal{A}_{\varepsilon}^{(n)}(X_r,W_{\mathit{In}(r)})$$
 and such that
$$(\phi_{r_{1}}(\hat{\mathbf{w}}_{r_{1}}),\cdots,\phi_{r_{n(r)}}(\hat{\mathbf{w}}_{r_{n(r)}}))=(\phi_{r_{1}}(\mathbf{w}_{r_{1}}),\cdots,\phi_{r_{n(r)}}(\mathbf{w}_{r_{n(r)}})).$$
 
\noindent \textit{Probability of error:} 
Before computing the error probability let us observe that if for all $u \in \mathcal{V} \setminus\{r\}$ message $\mathbf{w}_u$ is correctly encoded at vertex $u$ and correctly decoded at vertex $u_{out}$, the function $f(\mathbf{x}_{\mathcal{V}})$ can be computed with no error. To see this note first that at each step message $\mathbf{w}_u$ is chosen such that $(\mathbf{x}_u,\hat{\mathbf{w}}_{\mathit{In}(u)},\mathbf{w}_u)$ is jointly typical. Due to Claim c.ii. of Lemma~\ref{lem:jointlyTypicality}, this implies that $p(x_{u,i},w_{\mathit{In}(u),i},w_{u,i})>0$ for any $1 \leq i \leq n$. This together with \eqref{eq:treeAchieveSets} implies that $(x_{u,i},w_{\mathit{In}(u),i})\in w_{u,i}$, \textit{i.e.}, each component $w_{u,i}$ is an independent set in the graph  $$G_{X_{u},W_{\mathit{In}(u)}|X_{\mathit{Sup}(u)},W_{\mathit{Roots}(u)}}$$ that includes $(x_{u,i},w_{\mathit{In}(u),i})$. Moreover, due to the definition of conditional characteristic graph (Definition~\ref{def:condCharGraph}), by choosing the random variables $W_u$ recursively as in \eqref{eq:treeAchieveSets}, at each step  $$(W_u,W_{\mathit{Roots}(u)},X_{\mathit{Sup}(u)})$$
is sufficient for computing the function $f(X_{\mathcal{V}})$. Taking $u=O^{-1}(|\mathcal{V}|-1)$ implies that the root can compute the function by knowing $(W_{\mathit{In}(r)},X_r)$.

We now show that for any node $u\ne r$ the probability that message $\mathbf{w}_u$ is incorrectly encoded at vertex $u$ or incorrectly decoded at vertex $u_{out}$ can be made arbitrarily small by taking $n$ large enough. A union bound over the nodes then implies that the root can compute the function with arbitrarily high probability.

Equivalently, we show that the following two events happen with arbitrarily low probability. The first event happens when some of the (incoming) messages in $\mathbf{w}_{\mathit{In}(u)}$ are incorrectly decoded assuming that they all have been correctly encoded at nodes $\mathit{In}(u)$. The second event happens when message $\mathbf{w}_{u}$ is incorrectly encoded, \textit{i.e.}, 
when  no $\mathbf{w}_u$ is 
jointly  typical with $(\mathbf{x}_u,\hat{\mathbf{w}}_{\mathit{In}(u)})$.\footnote{ For leaves there is only the second event.}

The probability of the second event is negligible for $ n $ large enough due to the covering lemma (Lemma~\ref{lem:covering}) whenever
\begin{align}
R_{W_u} > I(X_u,W_{\mathit{In}(u)};W_u) + \delta(\varepsilon_i), \qquad i=O(u) \label{eq:sizeOfCodebooks}
\end{align}
where $\delta(\varepsilon_i)$ tends to zero whenever $\varepsilon_i$ goes to zero.

We now bound the probability of the first event assuming that the incoming neighbors correctly encoded their messages. By symmetry of the encoding and decoding
procedures, the probability of this event, averaged
over sources outcomes,  over $\mathbf{w}_v$'s, and over the 
binning assignments, is the same as the average
 probability conditioned on vertex $v$ correctly selecting $\mathbf{W}_v^{(1)}$, $v \in \mathit{In}(u)$. Hence, we compute the probability of the event 
\begin{align}
\lbrace \hat{\mathbf{W}}_{\mathit{In}(u)}\neq \mathbf{W}_{\mathit{In}(u)}^{(1)}\rbrace \label{eq:treeSecondEvent}
\end{align}
assuming that each vertex $v\in \mathit{In}(u)$ has previously selected $\mathbf{W}_v^{(1)}$ such that
\begin{align}
(\mathbf{W}_{\mathit{In}(v)},\mathbf{X}_v,\mathbf{W}_v^{(1)}) \in \mathcal{A}_{\varepsilon_{O(v)}}^{(n)}(W_{\mathit{In}(v)},X_v,W_v). \label{eq:proofAchievEncode}
\end{align}
Denote the elements of a set $\mathcal{S} \subseteq \mathit{In}(u)$ by $u_{s_1}$, $u_{s_2}$, $\cdots$, $u_{s_{|\mathcal{S}|}}$ and let $j_l$ be a natural number such that $$1 \leq j_l \leq 2^{nR_{W_{u_{l}}}}\quad \quad 1 \leq l \leq n(u)$$ where $\{u_{1},u_{2},\cdots,u_{n(u)}\}=\mathit{In}(u)$.

Define event $\EuScript{E}(j^{n(u)})$ as
\begin{align*}
\EuScript{E}(j^{n(u)})\defeq \{(\mathbf{W}^{(j^{n(u)})}_{\mathit{In}(u)},\mathbf{X}_u)&\in
\mathcal{A}_{\varepsilon_i}^{(n)}(W_{\mathit{In}(u)},X_u), \\
\phi_{u_{1}}(\mathbf{W}_{u_{1}}^{(j_1)})&=\phi_{u_{1}}(\mathbf{W}_{u_{1}}^{(1)}) \\
\phi_{u_{2}}(\mathbf{W}_{u_{2}}^{(j_2)})&=\phi_{u_{2}}(\mathbf{W}_{u_{2}}^{(1)}) \\
&\hspace{-.05 cm}\cdots \\
\phi_{u_{n(u)}}(\mathbf{W}_{u_{n(u)}}^{(j_{n(u)})})&=\phi_{u_{n(u)}}(\mathbf{W}_{u_{n(u)}}^{(1)})\} \\
\end{align*}
where 
$$\mathbf{W}^{(j^{n(u)})}_{\mathit{In}(u)}\defeq(\mathbf{W}^{(j_1)}_{u_{1}},\mathbf{W}^{(j_2)}_{u_{2}},\cdots,\mathbf{W}^{(j_{n(u)})}_{u_{n(u)}}).$$

The probability of the event \eqref{eq:treeSecondEvent} is upper bounded as
\begin{align} 
P(\hat{\mathbf{W}}_{\mathit{In}(u)}\neq \mathbf{W}_{\mathit{In}(u)}^{(1)})&=P\Big( \EuScript{E}^c((1,1,\cdots,1)\bigcup\big(\bigcup \limits_{j^{n(u)}\ne (1,1,\cdots,1)}\EuScript{E}(j^{n(u)})\big)\Big)\nonumber\\[.5 cm]
&=P\Bigg(\EuScript{E}^c((1,1,\cdots,1))\bigcup\Big( \bigcup \limits_{\substack {  \mathcal{S}:\\   \varnothing \neq \mathcal{S} \subseteq \mathit{In}(u)}} \bigcup \limits_{\substack{j^{n(u)}:j_{\mathcal{S}'}=(1,1,\cdots,1),\\j_{s_1}\neq 1,j_{s_2}\neq 1, \cdots ,j_{s_{|\mathcal{S}|}}\neq 1}}\EuScript{E}(j^{n(u)})\Big)\Bigg) \nonumber\\
&\leq P(\EuScript{E}^c((1,1,\cdots,1))) + \sum \limits_{\substack{\mathcal{S}:\\   \varnothing \neq \mathcal{S} \subseteq \mathit{In}(u)}} \sum \limits_{\substack{j^{n(u)}:\\j_{s_1}\neq 1\\j_{s_2}\neq 1\\ \cdots \\j_{s_{|\mathcal{S}|}}\neq 1\\j_{\mathcal{S}'}=(1,1,\cdots,1)}}P(\EuScript{E}{(j^{n(u)})}), \label{eq:errorsum}
\end{align}
where $\mathcal{S}'=\mathit{In}(u)\setminus \mathcal{S}$.

We bound each of the two terms on the right-hand side of \eqref{eq:errorsum}. For the first term, according to \eqref{eq:proofAchievEncode} and the properties of jointly 
 typical sequences (Lemmas~\ref{lem:jointlyTypicality}, \ref{lem:conditionalTypicalaity}, \ref{lem:markovTypicality}, and \ref{lem:independentTypicalaity}), we have
\begin{align} 
P(\EuScript{E}^c{(1,1,\cdots,1)}) \leq \delta(\varepsilon_i). \nonumber 
\end{align}
where $\delta(\varepsilon_i)\overset{\varepsilon_i\to 0}{\longrightarrow}0$.

Now for the second term. For any $\mathcal{S}$ such that $\varnothing \neq \mathcal{S} \subseteq \mathit{In}(u)$, $\mathcal{S}'=\mathit{In}(u)\setminus \mathcal{S}$, and any $j^{n(u)}$ such that $$j_{s_1}\neq 1, j_{s_2}\neq 1,\cdots, j_{s_{|\mathcal{S}|}}\neq 1$$ and $$j_{\mathcal{S}'}=(1,1,\cdots,1)$$ we have
 \begin{align} 
 P(\EuScript{E}{(j^{n(u)})}) \leq &2^{-n\sum \limits_{v \in \mathcal{S}}R_{v}}\cdot (\prod \limits_{i=1}^{|\mathcal{S}|-1}  2^{-n(I(W_{u_{s_1}},\cdots,W_{u_{s_i}};W_{u_{s_{i+1}}})-\delta_i(\varepsilon_i))}) \cdot  2^{-n(I(W_{\mathcal{S}};X_u,W_{\mathcal{S}'})-\delta_{|\mathcal{S}|}(\varepsilon_i))}. \label{eq:probErrorJnu}
\end{align} 
Since 
 \begin{align} 
 \sum \limits_{\substack{j^{n(u)}:\\j_{s_1}\neq 1\\j_{s_2}\neq 1\\ \cdots \\j_{s_{|\mathcal{S}|}}\neq 1\\j_{\mathcal{S}'}=(1,1,\cdots,1)}} 1&\leq
 \prod \limits_{i=1}^{|\mathcal{S}|} 2^{nR_{W_{u_{s_i}}}}, \nonumber 
 \end{align}
by using \eqref{eq:sizeOfCodebooks} and \eqref{eq:probErrorJnu} we conclude that the second term on the right-hand side of \eqref{eq:errorsum} is negligible for $n$  large enough provided that\footnote{Note that the summation over the sets $\mathcal{S}$ in the second term on the right-hand side of \eqref{eq:errorsum} involves a constant number of elements  that does not depend on $n$.}
\begin{align*}
\sum \limits_{v \in \mathcal{S}}R_{v}&>\sum \limits_{i=1}^{|\mathcal{S}|}
I(X_{u_{s_i}},W_{\mathit{In}(u_{s_i})};W_{u_{s_i}})-\sum \limits_{i=1}^{|\mathcal{S}|-1}  I(W_{u_{s_1}},\cdots,W_{u_{s_i}};W_{u_{s_{i+1}}})-I(W_{\mathcal{S}};X_u,W_{\mathcal{S}'})\\
&=\sum \limits_{i=1}^{|\mathcal{S}|}
I(X_{u_{s_i}},W_{\mathit{In}(u_{s_i})};W_{u_{s_i}})-\sum \limits_{i=1}^{|\mathcal{S}|-1}  \sum \limits_{j=1}^{i} I(W_{u_{s_j}};W_{u_{s_{i+1}}}|W_{u_{s_{j+1}}},\cdots,W_{u_{s_i}})\\
&\hspace{5.8 cm}- \sum \limits_{j=1}^{|\mathcal{S}|} I(W_{u_{s_j}};X_u,W_{\mathcal{S}'}|W_{u_{s_{j+1}}},\cdots,W_{u_{s_{|\mathcal{S}|}}})\\
&=\sum \limits_{i=1}^{|\mathcal{S}|}
I(X_{u_{s_i}},W_{\mathit{In}(u_{s_i})};W_{u_{s_i}})-\sum \limits_{j=1}^{|\mathcal{S}|-1}  \sum \limits_{i=j+1}^{|\mathcal{S}|} I(W_{u_{s_j}};W_{u_{s_{i}}}|W_{u_{s_{j+1}}},\cdots,W_{u_{s_{i-1}}})\\
&\hspace{5.8 cm}- \sum \limits_{j=1}^{|\mathcal{S}|} I(W_{u_{s_j}};X_u,W_{\mathcal{S}'}|W_{u_{s_{j+1}}},\cdots,W_{u_{s_{|\mathcal{S}|}}})\\
&=\sum \limits_{i=1}^{|\mathcal{S}|}
I(X_{u_{s_i}},W_{\mathit{In}(u_{s_i})};W_{u_{s_i}})-\sum \limits_{j=1}^{|\mathcal{S}|-1}  I(W_{u_{s_j}};W_{u_{s_{j+1}}},\cdots,W_{u_{s_{|\mathcal{S}|}}})\\
&\hspace{5.8 cm}- \sum \limits_{j=1}^{|\mathcal{S}|} I(W_{u_{s_j}};X_u,W_{\mathcal{S}'}|W_{u_{s_{j+1}}},\cdots,W_{u_{s_{|\mathcal{S}|}}})\\
&=\sum \limits_{i=1}^{|\mathcal{S}|}
I(X_{u_{s_i}},W_{\mathit{In}(u_{s_i})};W_{u_{s_i}})-\sum \limits_{j=1}^{|\mathcal{S}|}  I(W_{u_{s_j}};W_{u_{s_{j+1}}},\cdots,W_{u_{s_{|\mathcal{S}|}}},X_u,W_{\mathcal{S}'})\\
&=\sum \limits_{i=1}^{|\mathcal{S}|}
I(X_{u_{s_i}},W_{\mathit{In}(u_{s_i})};W_{u_{s_i}})-  I(W_{u_{s_i}};W_{u_{s_{i+1}}},\cdots,W_{u_{s_{|\mathcal{S}|}}},X_u,W_{\mathcal{S}'})\\
&=\sum \limits_{i=1}^{|\mathcal{S}|}
H(W_{u_{s_i}}|W_{u_{s_{i+1}}},\cdots,W_{u_{s_{|\mathcal{S}|}}},X_u,W_{\mathcal{S}'})-H(W_{u_{s_i}}|X_{u_{s_i}},W_{\mathit{In}(u_{s_i})}) \end{align*}
\begin{align*}
&\stackrel{(a)}{=}\sum \limits_{i=1}^{|\mathcal{S}|}
I(X_{\mathcal{S}},W_{\mathit{In}(\mathcal{S})};W_{u_{s_i}}|W_{u_{s_{i+1}}},\cdots,W_{u_{s_{|\mathcal{S}|}}},X_u,W_{\mathcal{S}'})\\
&=I(X_{\mathcal{S}},W_{\mathit{In}(\mathcal{S})};W_{\mathcal{S}}|X_u,W_{\mathcal{S}'}) \nonumber
\end{align*}
where $(a)$ holds due to Markov chains \eqref{eq:treeAchieveMarkovs2}. This completes the  achievability part of the Theorem.

 It remains to show that different orderings yield the same achievable regions. For this it is sufficient to establish the following claim whose proof is deferred to Appendix~\ref{subSec:ProofOfequivalentOrdering}. 
 
 \begin{claim} \label{claim:ClaimWStrange} If $W_{\mathcal{V}\setminus \{r\}}$ satisfies conditions \eqref{eq:treeAchieveMarkovs2} and \eqref{eq:treeAchieveSets} for an ordering $O$, then $W_{\mathcal{V}\setminus \{r\}}$ also satisfies these conditions for any other ordering $O'$.  \hfill $\tiny{\square}$
\end{claim} 
This  completes the proof or the theorem.
\end{IEEEproof} 

\begin{IEEEproof}[Proof of Theorem~\ref{th:treeRate}]
Suppose that random variables $X_{\mathcal{V}}$ satisfy the Markov property (Definition~\ref{def:MarkovProperty}). We show that the inner bound in Theorem~\ref{th:treeAchieve} is tight with an outer bound derived from the outer bound of Theorem~\ref{th:treeUpper}. Without loss of generality, we suppose that the set of vertices are $\{1,2,\cdots,m\}$ and that the ordering is the natural ordering given by $O(u)=u$, for $1 \leq u \leq m$, with $r=m$.

\subsection*{Outer bound}
 Consider the following constraints in Theorem~\ref{th:treeUpper}
\begin{align}
R_{u} \geq H(G_{X_{{\mathit{Child}(u)}}|X_{\mathit{Child}(u)^c}}) \hspace{1 cm} u \in  \mathcal{V}\setminus \{r\} \label{eq:treeOuterWeak}
\end{align} 
which are derived by letting $\mathcal{S}=\mathit{Child}(u)$. Considering only these constraints gives a weaker outer bound than the one of Theorem~\ref{th:treeUpper}. 

\subsection*{Inner bound} We show that \eqref{eq:treeOuterWeak} is achievable using Theorem~\ref{th:treeAchieve}, thereby completing the proof of the theorem. This is done in a number of steps. We first simplify the rate constraints \eqref{eq:treeAchieveRates} in Theorem~\ref{th:treeAchieve} using the following claim whose proof is deferred to Appendix~\ref{subSec:treeIndependence}. Then, we show that using these simplified rate constraints yield \eqref{eq:treeOuterWeak}.
 \begin{claim} \label{claim:treeIndependence}
Suppose that the random variables $X_{\mathcal{V}}$ satisfy the Markov property and that the random variables $W_{\mathcal{V}\setminus\{r\}}$ satisfy the Markov chain conditions \eqref{eq:treeAchieveMarkovs}. Then, the set of pairs of random variables 
\begin{align*}
((X_{\mathit{Child}(u_{1})},W_{\mathit{Child}(u_{1})}),\cdots,(X_{\mathit{Child}(u_{n(u)})},W_{\mathit{Child}(u_{n(u)})}),(X_{\mathit{Child}(u)^c},W_{\mathit{Sub}(u)\setminus\mathit{Child}(u)}))
\end{align*}
are jointly independent given $X_u$ for $u \in \mathcal{V}$, where $\{u_{1},\cdots, u_{n(u)}\}=\mathit{In}(u).$
In particular, this implies that the pair $$(X_{\mathit{Child}(u) \setminus \{u\}},W_{\mathit{Child}(u)\setminus \{u\}})$$ is independent of the pair $$(X_{\mathit{Child}(u)^c},W_{\mathit{Sub}(u)\setminus \mathit{Child}(u)})$$ given the value of $X_u$, for any $u \in \mathcal{V}\setminus \{r\}$. \hfill $\square$
\end{claim}

Consider the rate constraints \eqref{eq:treeAchieveRates} in Theorem~\ref{th:treeAchieve}. Claim \ref{claim:treeIndependence} implies that for the terms on the right-hand side of  \eqref{eq:treeAchieveRates} we have
\begin{align}
I(X_{\mathcal{S}},W_{\mathit{In}(\mathcal{S})};W_{\mathcal{S}}|X_{u},W_{\mathcal{S}'})=\sum \limits_{v \in \mathcal{S}} I(X_{v},W_{\mathit{In}(v)};W_{v}|X_{u}). \label{eq:treeRatesProofs1}
\end{align} 
Hence, the rate constraints \eqref{eq:treeAchieveRates} reduce to the following constraints:
\begin{align}
 R_u\geq I(X_{u},W_{\mathit{In}(u)};W_u|X_{u_{out}}) \hspace{1 cm} u \in  \mathcal{V}\setminus \{r\} \label{eq:treeRatesProofs2}.
 \end{align}  
 Therefore, we may consider the constraints \eqref{eq:treeRatesProofs2} instead of \eqref{eq:treeAchieveRates}. Moreover, using Remark~\ref{remark:ProofOfClaimMarkovStrange} (p.\pageref{remark:ProofOfClaimMarkovStrange}), we consider Markov chains \eqref{eq:treeAchieveMarkovs} instead of Markov chains \eqref{eq:treeAchieveMarkovs2}. 
 
 \subsection*{Inner and outer bound match: induction}
We  now show that the above inner bound matches the outer bound \eqref{eq:treeOuterWeak}.  For this, it is sufficient to show that
\begin{align}
I(X_{u},W_{\mathit{In}(u)};W_u|X_{u_{out}}) \leq H(G_{X_{{\mathit{Child}(u)}}|X_{\mathit{Child}(u)^c}}) \hspace{1 cm} u \in  \mathcal{V}\setminus \{r\} \label{eq:treeInnerOuter}
\end{align}
for a specific choice of $W_{\mathcal{V}\setminus \{r\}}$ that satisfy the constraints \eqref{eq:treeAchieveSets} in Theorem~\ref{th:treeAchieve} and \eqref{eq:treeAchieveMarkovs} in Remark~\ref{remark:ProofOfClaimMarkovStrange}. Rewrite inequalities \eqref{eq:treeInnerOuter} as
\begin{align}\label{Milad}
I(X_{1},W_{\mathit{In}(1)};W_1|X_{1_{out}}) &\leq H(G_{X_{{\mathit{Child}(1)}}|X_{\mathit{Child}(1)^c}})\notag \\
I(X_{2},W_{\mathit{In}(2)};W_2|X_{2_{out}}) &\leq H(G_{X_{{\mathit{Child}(2)}}|X_{\mathit{Child}(2)^c}}) \notag\\
& \cdots \notag \\
I(X_{m},W_{\mathit{In}(m)};W_m|X_{m_{out}}) &\leq H(G_{X_{{\mathit{Child}(m)}}|X_{\mathit{Child}(m)^c}}). 
\end{align}
Note that the first $u-1$ inequalities do not depend on $W_u$, for $1 \leq u \leq m$.
 Using induction, we show that for any $1 \leq k \leq m$, the first $k$ inequalities hold for some $W_1^*, \cdots, W_k^*$ that satisfy conditions \eqref{eq:treeAchieveSets} and \eqref{eq:treeAchieveMarkovs}, and such that $W_u^{*}$, $1 \leq u \leq k$, takes values only over maximal independent sets of
 \begin{align}\label{eq:gx}G_{X_{u},W^*_{\mathit{In}(u)}|X_{\mathit{Sup}(u)},W^*_{\mathit{Roots}(u)}}.
 \end{align}
 \begin{itemize}
\item \textit{Induction base:} For $k=1$, we have $\mathit{In}(1)=\varnothing$ and $\mathit{Child}(1)=\{1\}$. Moreover, we have $I(X_{1};W_1|X_{1_{out}})=I(X_{1};W_1|X_{2}^m)$ due to   Claim~\ref{claim:treeIndependence}. Hence, to show that the first inequality in \eqref{Milad} holds it suffices to show that there exists $W_1$ such that $$I(X_{1};W_1|X_{2}^m) \leq H(G_{X_{1}|X_2^m}).$$ 
A natural choice is to pick $W_1=W_1^{*}$ as the random variable that achieves
$H(G_{X_{1}|X_2^m})$, {\it{i.e.}}, the one that minimizes
$$I(X_{1};W_1|X_{2}^m),$$ 
among all $W_1$'s such that  
\begin{align}
X_{1} &\in W_1  \in 
\Gamma^*(G_{X_{1}|X_{2}^m}) \nonumber \\
&W_1-X_{1}-X_{2}^m. \nonumber
\end{align}
Trivially conditions \eqref{eq:treeAchieveSets} and \eqref{eq:treeAchieveMarkovs} are satisfied by $W^*_1$. Since $\Gamma^*(G_{X_{1}|X_{2}^m})$ corresponds to the maximal independent sets of the conditional characteristic graph \eqref{eq:gx}, the case $k=1$ is proved.

\item \textit{Induction step:} Suppose that the first $k-1$ inequalities in \eqref{Milad} hold for some $W_1^*, \cdots, W_{k-1}^*$ that satisfy conditions \eqref{eq:treeAchieveSets} and \eqref{eq:treeAchieveMarkovs}, and such that $W_u^{*}$, $1 \leq u \leq k-1$, take values over the maximal independent sets of $$G_{X_{u},W^*_{\mathit{In}(u)}|X_{\mathit{Sup}(u)},W^*_{\mathit{Roots}(u)}}.$$
 
 We now show how to choose a proper $W_k^*$ such that the $k$-th inequality holds. Note that random variable $W_k$ does not appear in the first $k-1$ inequalities (however, some of the $W_i$, $i<k$, appear in the $k$\/th inequality).
 
The following claim, whose proof is deferred to Appendix~\ref{subSec:treeGraphEntropy}, says that the graph entropy term on the right-hand side of the $k$-th inequality in \eqref{Milad} is equal to another graph entropy that we shall analyze here below:
\begin{claim} \label{claim:treeGraphEntropy}
Suppose that the random variables $X_{\mathcal{V}}$ satisfy the Markov property and that the random variables $W_{\mathcal{V}\setminus\{r\}}$ satisfy the conditions \eqref{eq:treeAchieveSets} and \eqref{eq:treeAchieveMarkovs}. Then,
\begin{align}
H(G_{X_{{\mathit{Child}(k)}}|X_{\mathit{Child}(k)^c}})=H(G_{X_{\mathit{Child}(k)}|X_{\mathit{Sup}(k)},W_{\mathit{Roots}(k)}}).
\end{align}
\hfill $\square$
\end{claim}
Using this claim, the $k$-th inequality becomes 
\begin{align}
I(X_{k},W^*_{\mathit{In}(k)};W_k|X_{k_{out}}) &\leq H(G_{X_{\mathit{Child}(k)}|X_{\mathit{Sup}(k)},W^*_{\mathit{Roots}(k)}}).  \label{eq:equivalentKthInequality}
\end{align}
We show that this inequality holds for a proper choice of $W_k$ which completes the proof of the induction step, and hence the proof of the tightness of the inner and the outer bounds under the Markov property. 

In the remaining of the proof we first introduce a random variable $W_k^{'}$ which satisfies the $k$\/th inequality and condition \eqref{eq:treeAchieveMarkovs}. Then, by a change of alphabet we define $W^*_k$ which, in addition, takes values over the maximal independent sets of $$G_{X_{k},W^*_{\mathit{In}(k)}|X_{\mathit{Sup}(k)},W^*_{\mathit{Roots}(k)}}$$ and satisfies condition~\eqref{eq:treeAchieveSets}. This shall complete the proof of the induction step and thereby conclude the proof of the theorem.
\setlength{\IEEEilabelindent}{0pt} 
\begin{itemize}
\item \textit{Defining $W_k^{'}$:}
Let $W_k^{'}$ be the random variable that achieves 
$H(G_{X_{\mathit{Child}(k)}|X_{\mathit{Sup}(k)},W^*_{\mathit{Roots}(k)}})$, {\it{i.e.}}, the one that minimizes
$$I(X_{\mathit{Child}(k)};W|X_{\mathit{Sup}(k)},W^{*}_{\mathit{Roots}(k)})$$ 
among all $W$'s such that  
\begin{align}
&  X_{\mathit{Child}(k)} \in W  \in 
\Gamma^*(G_{X_{\mathit{Child}(k)}|X_{\mathit{Sup}(k)},W^{*}_{\mathit{Roots}(k)}}) \nonumber \\
&\hspace{.8 cm} W-X_{\mathit{Child}(k)}-(X_{\mathit{Sup}(k)},W^{*}_{\mathit{Roots}(k)}). \label{treeMark1}
\end{align}
Suppose that $(W^{'}_k,X_{\mathcal{V}},W^{*}_{\mathit{Roots}(k)})$ and $(X_{\mathcal{V}},W^{*}_{\mathit{Sub} (k)})$ are distributed according to some joint distribution 
\begin{align}\label{marg1}
 p_{(W^{'}_k,X_{\mathcal{V}},W^{*}_{\mathit{Roots}(k)})}(\cdot,\cdot,\cdot)
 \end{align}
  and  
  \begin{align}\label{marg2}
  p_{(X_{\mathcal{V}},W^{*}_{\mathit{Sub} (k)})}(\cdot,\cdot),
  \end{align}
  respectively, where the latter distribution is defined through the induction assumption. Note that, by definition, $\mathit{Roots}(k)\subseteq \mathit{Sub} (k)$, hence $W^{*}_{\mathit{Roots}(k)}$ involves a subset of the random variables $W^{*}_{\mathit{Sub} (k)}$.
  
  Now define the joint distribution of $(W^{'}_k,X_{\mathcal{V}},W^{*}_{\mathit{Sub} (k)})$ as 
\begin{align}\label{Milad2}
p_{(W^{'}_k,X_{\mathcal{V}},W^{*}_{\mathit{Sub} (k)})}(w^{'}_k,x_{\mathcal{V}},w^{*}_{\mathit{Sub} (k)})\defeq
 p_{(X_{\mathcal{V}},W^{*}_{\mathit{Sub} (k)})}(x_{\mathcal{V}},w^{*}_{\mathit{Sub} (k)}) \cdot p_{(W^{'}_k|X_{\mathit{Child}(k)})}(w^{'}_k|x_{\mathit{Child}(k)}).
\end{align}
Note that this distribution keeps the marginals \eqref{marg1} (due to the Markov chain \eqref{treeMark1}) and \eqref{marg2}. Moreover, Definition~\eqref{Milad2} yields the following Markov chains
\begin{align*}
&W^{'}_k-X_{\mathit{Child}(k)}-(X_{\mathit{Child}(k)^c},W^{*}_{\mathit{Sub}(k)})\\
W_k^{'}-&(X_k,W^{*}_{\mathit{In}(k)})-(X_{\mathit{Child}(k)^c },W^{*}_{\mathit{Sub}(k)\setminus \mathit{Child}(k)}),
\end{align*}
where the second Markov chain holds because of Claim~\ref{claim:WuisinWAina} whose proof is deferred to Appendix~\ref{subSec:WuisinWAina}. These Markov chains imply that  inequality \eqref{eq:equivalentKthInequality} holds, {\it{i.e.}}
\begin{align}\label{treeHyp}
I(X_k,W^{*}_{\mathit{In}(k)};W^{'}_k|X_{\mathit{Sup}(k)},W^{*}_{\mathit{Roots}(k)})\leq
I(X_{\mathit{Child}(k)};W^{'}_k|X_{\mathit{Sup}(k)},W^{*}_{\mathit{Roots}(k)}).
\end{align}

\begin{claim} \label{claim:WuisinWAina} 
Condition \eqref{eq:treeAchieveMarkovs} holds for $W_k=W^{'}_k$. \hfill $\square$
\end{claim}

\item \textit{Defining $W^*_k$ from $W^{'}_k$ :} For $w^{'}_k\in W^{'}_k$ define
\begin{align*}
\mathcal{B}_{w^{'}_k} \defeq \{(w_{k_{1}},\cdots,w_{k_{n(k)}},x_k)|&(w_{k_{1}},\cdots,w_{k_{n(k)}},x_k) \in (\mathcal{W}^*_{k_{1}},\cdots,\mathcal{W}^*_{k_{n(k)}},\mathcal{X}_k), \\ &\exists x_{\mathit{Child}(k)} \in  w_k':  p(x_{\mathit{Child}(k_{i})},w_{k_{i}})>0, \forall i \in \{1,2,\ldots,n(k)\}\}
\end{align*}
where $\{k_{1},\cdots,k_{n(k)}\}=\mathit{In}(k)$ and $\mathcal{W}^*_{k_{i}} \subseteq \Gamma^*(G_{X_{k_i},W^*_{\mathit{In}(k_i)}|X_{\mathit{Sup}(k_i)},W^*_{\mathit{Roots}(k_i)}})$.

First we show that $w_k^{'}$ and $\mathcal{B}_{w^{'}_k}$ are in one-to-one correspondence, \textit{i.e.}, there is no $w_1,w_2 \in \mathcal{W}^{'}_k$ with $w_1 \neq w_2$ such that $\mathcal{B}_{w_1}=\mathcal{B}_{w_2}$.  This can be deduced from the following claim whose proof is deferred to Appendix~\ref{subSec:treeEquiv}. 

\begin{claim} \label{claim:treeEquiv}
If $(w_{k_{1}},\cdots,w_{k_{n(k)}},x_k) \in \mathcal{B}_{w}$, $w\in \mathcal{W}^{'}_k$, and 
$p(x_{\mathit{Child}(k_{i})},w_{k_{i}})>0$, $1 \leq  i \leq n(k)$, then $$x_{\mathit{Child}(k)}=(x_{\mathit{Child}(k_{1})},\cdots,x_{\mathit{Child}(k_{n(k)})},x_k) \in w.$$\hfill $\square$
\end{claim} 
By this claim one can verify that if $\mathcal{B}_{w_1}=\mathcal{B}_{w_2}$ for some $w_1,w_2 \in \mathcal{W}^{'}_k$, then $w_1 = w_2$, which shows the  one-to-one correspondence between  $w_k^{'}$ and $\mathcal{B}_{w^{'}_k}$.
 
Let random variable $W^*_k$ take values over the set
$$\mathcal{W}^*_k\defeq \{\mathcal{B}_{w^{'}_k}:w^{'}_k \in \mathcal{W}^{'}_k \}$$ 
with conditional distribution 
\begin{align}
p(w^*_k|x_{\mathcal{V}},w^{*}_{\mathit{Sub}(k)})=p(w^{*}_k|x_{\mathit{Child}(k)})=p(w^{'}_k|x_{\mathit{Child}(k)}), \nonumber
\end{align}
where $w^{'}_k \in \mathcal{W}^{'}_k$  is the unique value such that $\mathcal{B}_{w^{'}_k}=w^{*}_k$.
\item \textit{Showing that the $k$-th inequality holds:} We first show that $W^*_k$ satisfies conditions \eqref{eq:treeAchieveSets} and \eqref{eq:treeAchieveMarkovs} and that $W^*_k$ takes values only over maximal independent sets. This can be deduced from parts a. and b. of the following claim whose proof is deferred to Appendix~\ref{subSec:WuisinWAin}. 
\begin{claim} \label{claim:WuisinWAin} We have the following relations for $W^*_k$:
\begin{itemize}
\item[a. ] The Markov chain $$W_k^{*}-(X_k,W^{*}_{\mathit{In}(k)})-(X_{\mathit{Child}(k)^c },W^{*}_{\mathit{Sub}(k)\setminus \mathit{Child}(k)})$$ 
 holds (equivalently, condition~\eqref{eq:treeAchieveMarkovs} holds);
\item[b. ]   $$(X_k,W^{*}_{\mathit{In}(k)}) \in W^*_k \in \Gamma^*(G_{X_k,W^{*}_{\mathit{In}(k)}|X_{\mathit{Sup}(k)},W^{*}_{\mathit{Roots}(k)}})\,;$$
\item[c. ] \begin{align*}
I(X_k,W^{*}_{\mathit{In}(k)};W^{*}_k|X_{\mathit{Sup}(k)},W^{*}_{\mathit{Roots}(k)})=I(X_k,W^{*}_{\mathit{In}(k)};W^{'}_k|X_{\mathit{Sup}(k)},W^{*}_{\mathit{Roots}(k)}).
\end{align*}
\end{itemize}
\hfill $\square$
\end{claim}
Claim~\ref{claim:WuisinWAin}.c, together with \eqref{treeHyp}, and the fact that 
\begin{align}
I(X_{\mathit{Child}(k)};W^{'}_k|X_{\mathit{Sup}(k)},W^{*}_{\mathit{Roots}(k)})=H(G_{X_{\mathit{Child}(k)}|X_{\mathit{Sup}(k)},W^{*}_{\mathit{Roots}(k)}}) \nonumber
\end{align}
implies that  the $k$-th inequality holds. This completes the induction step. 
\end{itemize}
\end{itemize}
\end{IEEEproof}
\begin{IEEEproof}[Proof of Corollary~\ref{th:cor}]
From Theorem~\ref{th:treeRate} we have
\begin{align*}
R_{u} &\geq H(G_{X_{{\mathit{Child}(u)}}|X_{\mathit{Child}(u)^c}})\\
&=I(W_u^*;X_{{\mathit{Child}(u)}}|X_{\mathit{Child}(u)^c}) \\
&=H(W_u^*|X_{\mathit{Child}(u)^c})-H(W_u^*|X_{\mathcal{V}})  \\
&\stackrel{(a)}{=}H(W_u^*)-H(W_u^*|X_{\mathcal{V}})   \\
&\stackrel{(b)}{=}H(W_u^*)
\end{align*}
where $(a)$ follows from the independence of the sources, the Markov chains \eqref{eq:treeAchieveMarkovs2}, and Claim~\ref{claim:treeIndependence} stated in the proof of Theorem~\ref{th:treeRate}, and where $(b)$ follows from Lemma~\ref{lem:condGraphIndependent} since each vertex is included in exactly one maximal independent set. 
\end{IEEEproof}

{\small{
\bibliographystyle{plain}
\bibliography{bib/bib}}} 
\appendix
\subsection{Proof of Lemma~\ref{lem:condGraphIndependent}} \label{app:condGraphIndependent}
Each vertex is contained in at least one maximal independent set. Suppose, by way of contradiction, that there exists a vertex  $l \in \mathcal{L}$ that belongs to two maximal independent sets $w_1,w_2 \in \Gamma^*(G_{L|K})$. This means that there exist some $l_1 \in w_1$ and $l_2 \in w_2$ such that $l_1$ and $l_2$ are connected in $G_{L|K}$, \textit{i.e.}, there exist some $s_1,s'_1\in \mathcal{S}_1$, $s_2,s'_2 \in \mathcal{S}_2$ and $k \in \mathcal{K}$ such that
\begin{align}
p(s_1,s_2,l_1,k)\cdot p(s'_1,s'_2,l_2,k)&=p(s_1,l_1)\cdot p(s'_1,l_2)\cdot p(k,s_2)\cdot p(k,s'_2)>0 \label{eq:prelimIndependence1}	\\
f(s_1,s_2)&\neq f(s'_1,s'_2).  \label{eq:prelimIndependence2}
\end{align}
Now take any $s''_1 \in \mathcal{S}_1$ such that $p(s''_1,l)>0$. This, together with \eqref{eq:prelimIndependence1}, and the fact that both vertex pairs $(l_1,l)$ and $(l_2,l)$ are  disconnected in $G_{L|K}$ implies that
 $$f(s_1,s_2)=f(s''_1,s_2)=f(s'_1,s'_2),$$
 which contradicts \eqref{eq:prelimIndependence2}.

\subsection{Proof of Remark~\ref{remark:ProofOfClaimMarkovStrange}} \label{subSec:ProofOfClaimMarkovStrange}
Note that the Markov chains \eqref{eq:treeAchieveMarkovs2} trivially imply the Markov chains \eqref{eq:treeAchieveMarkovs} since the set of random variables $W_{\mathit{Sub}_O(u) \setminus \mathit{Child}(u)}$ is contained in the set of random variables $W_{\mathit{Strangers}(u)}$. 

We first show the reverse implication through the example in Fig.~\ref{fig:rootedDirectedTree2} with the natural ordering given by the labels of the nodes. We show the implication for vertex $5$. For this vertex the Markov chain \eqref{eq:treeAchieveMarkovs} becomes
\begin{align}
W_5-(X_5,W_1^2)-(X_3^4,X_6^{10},W_3^4). \label{eq:MarkovEquivalent1}
\end{align}
Also, for vertex $6$ the Markov chain \eqref{eq:treeAchieveMarkovs} corresponds to
\begin{align}
W_6-X_6-(X_1^5,X_7^{10},W_1^5). \label{eq:MarkovEquivalent3}
\end{align}
Combining \eqref{eq:MarkovEquivalent1} and \eqref{eq:MarkovEquivalent3} yields the Markov chain\footnote{Notice that random variables $A,B,C,D$ satisfy $A-B-(C,D)$ if and only if $A-B-C$ and $A-(B,C)-D$ hold.}
\begin{align*}
W_5-(X_5,W_1^2)-(X_3^4,X_6^{10},W_3^4,W_6).
\end{align*} 
 
Similarly, from this Markov chain and the corresponding Markov chains for vertices $7$ and $9$ in \eqref{eq:treeAchieveMarkovs} we get \begin{align}
W_5-(X_5,W_1^2)-(X_3^4,X_6^{10},W_3^4,W_6,W_7,W_9) \label{eq:MarkovEquivalent2}
\end{align} 
which corresponds to \eqref{eq:treeAchieveMarkovs2} with $u=5$.

In general, to show that the Markov chains \eqref{eq:treeAchieveMarkovs2}
hold, we observe that  \eqref{eq:treeAchieveMarkovs} and  \eqref{eq:treeAchieveMarkovs2} have the generic forms
\begin{align}\label{abc}
A-B-C
\end{align}
and 
\begin{align}\label{abcd}
A-B-(C,W_{d_1},W_{d_2},\ldots,W_{d_q})
\end{align}
respectively, where $$\{d_1,d_2,\ldots,d_q\}=\mathit{Strangers}(u) \setminus \mathit{Sub}(u)$$
and where, without loss of generality, the ordering is such that $$O(d_1)< O(d_2)<\ldots<O(d_q). $$
 To show that \eqref{eq:treeAchieveMarkovs} implies \eqref{eq:treeAchieveMarkovs2} one first shows that
\begin{align}\label{abcd1}
A-B-(C,W_{d_1})
\end{align} holds by using \eqref{abc} and  \eqref{eq:treeAchieveMarkovs} for the vertex $d_1$---in the example above $d_1=6$. Then one shows that
\begin{align}\label{abcd1d2}
A-B-(C,W_{d_1},W_{d_2})
\end{align}
holds using \eqref{abcd1} and \eqref{eq:treeAchieveMarkovs} for the vertex $d_2$---in the example above $d_2=7$. The argument is iterated for $d_3,\ldots,d_q$ thereby completing the proof.

\subsection{Proof of Claim~\ref{claim:ClaimWStrange}}
\label{subSec:ProofOfequivalentOrdering} 
As for the proof of Remark~\ref{remark:ProofOfClaimMarkovStrange}, consider first the particular network depicted in Fig.\ref{fig:rootedDirectedTree2} and let $O$ be the natural ordering given by the labels of the nodes and let $O'$ be obtained from $O$ by swapping the orders of the vertices $1$ and $2$, {\it{i.e.}} $$O'(1)=O(2)=2, O'(2)=O(1)=1,O'(i)=O(i), i\in\{3,4,\ldots,10\}.$$
We need to show that
\begin{align*}
X_1 &\in W_1 \in \text{M} (\Gamma(G_{X_1|W_2,X_3^{10}})) \\
X_2 &\in W_2 \in \text{M} (\Gamma(G_{X_2|X_1,X_3^{10}}))
\end{align*}
holds assuming that $W_1^9$ satisfy \eqref{eq:treeAchieveSets} and \eqref{eq:treeAchieveMarkovs} for ordering $O$. 

Since $W_1^9$ satisfy \eqref{eq:treeAchieveMarkovs2}, we have
\begin{align}
&W_1-X_1-(X_2^{10},W_2^4,W_6^7,W_9) \label{eq:SetEquivalenceMarkov1}\\
W_2&-X_2-(X_1,X_3^{10},W_1,W_3^4,W_6^7,W_9) \label{eq:SetEquivalenceMarkov2} 
\end{align}
and since $W_1^9$ satisfy \eqref{eq:treeAchieveSets} we have
\begin{align*}
X_1 &\in W_1 \in \text{M} (\Gamma(G_{X_1|X_2,X_3^{10}})) \\
X_2 &\in W_2 \in \text{M} (\Gamma(G_{X_2|W_1,X_3^{10}})).
\end{align*}
\begin{itemize}
\item To prove that $W_1 \in \text{M} (\Gamma(G_{X_1|W_2,X_3^{10}}))$, we need to show that for any $w_1 \in \mathcal{W}_1$, $x_1, x'_1 \in w_1$, $x_2, x'_2 \in \mathcal{X}_2$, $x_3^{10} \in X_3^{10}$, and $w_2 \in \mathcal{W}_2$ such that
\begin{align}
p(x_1,x_2,x_3^{10},w_2) \cdot p(x'_1,x'_2,x_3^{10},w_2)>0 \label{eq:SetEquivalenceProb2}
\end{align}
we have
\begin{align}
f(x_1,x_2,x_3^{10}) = f(x'_1,x'_2,x_3^{10}). \label{eq:SetEquivalenceFun2}
\end{align}
Note that \eqref{eq:SetEquivalenceProb2}, the fact that $x_1, x'_1 \in w_1$, and the Markov chain  \eqref{eq:SetEquivalenceMarkov1} imply that 
\begin{align*}
p(x_1,x_2,x_3^{10},w_1) \cdot p(x'_1,x'_2,x_3^{10},w_1)>0.
\end{align*}
This together with the facts that $x_2, x'_2 \in w_2$ (which can be deduced from $X_2 \in W_2$ and \eqref{eq:SetEquivalenceProb2}) and $W_2 \in \text{M} (\Gamma(G_{X_2|W_1,X_3^{10}}))$ implies \eqref{eq:SetEquivalenceFun2}.
\item To prove that $  W_2 \in \text{M} (\Gamma(G_{X_2|X_1,X_3^{10}})) $,  
we need to show that for any $ w_2 \in \mathcal{W}_2$,
$ x_2, x'_2 \in w_2 $, $ x_1 \in \mathcal{X}_1 $, and $x_3^{10} 
\in \mathcal{X}_3^{10}$ such that 
\begin{align}
p(x_1,x_2,x_3^{10}) \cdot p(x_1,x'_2,x_3^{10})>0, \label{eq:SetEquivalenceProb1}
\end{align}
we have 
\begin{align}
f(x_1,x_2,x_3^{10}) = f(x_1,x'_2,x_3^{10}). \label{eq:SetEquivalenceFun1}
\end{align}
Since $ P(X_1 \in W_1)=1 $, 
there exists $ w_1 \in W_1 $ such that $ p(w_1|x_1) >0 $. Then, using \eqref{eq:SetEquivalenceProb1} and Markov chain \eqref{eq:SetEquivalenceMarkov1} yields
\begin{align*}
p(x_1,x_2,x_3^{10},w_1) \cdot p(x_1,x'_2,x_3^{10},w_1)>0. 
\end{align*}
From the definition of $G_{X_2|W_1,X_3^{10}}$ we then deduce that equality \eqref{eq:SetEquivalenceFun1} holds.
\end{itemize}

In general, to show that $W_{\mathcal{V}\setminus \{r\}}$ satisfies condition  \eqref{eq:treeAchieveSets} for any ordering $O'$ it suffices to use the same arguments as above repeatedly. In more details, suppose $W_{\mathcal{V}\setminus \{r\}}$ satisfy  \eqref{eq:treeAchieveSets} for an ordering $O$ (over some given tree). Observe that any $O'$ can be obtained from $O$ by a sequence of neighbors' swaps (transpositions)---in the above example $O'$ is obtained from $O$ with one swap.  To show that \eqref{eq:treeAchieveSets} also holds for an ordering $O'$ one repeats the same arguments as above over the sequence of neighbor swaps that brings $O$ to $O'$. This completes the proof.

\subsection{Proof of Claim~\ref{claim:treeIndependence}}
\label{subSec:treeIndependence}

For notational simplicity, for a given a set $\mathcal{S}$ define $XW_{\mathcal{S}}\defeq (X_{\mathcal{S}},W_{\mathcal{S}})$ and $xw_{\mathcal{S}}\defeq (x_{\mathcal{S}},w_{\mathcal{S}})$.  To prove the claim, we show that the Markov chain
\begin{align}
W_{\mathit{Child}(u)}-X_{\mathit{Child}(u)}-(X_{\mathit{Child}(u)^c},W_{\mathit{Sub}(u)\setminus\mathit{Child}(u)}) \label{eq:claimIndMarkov1}
\end{align} 
holds for any vertex $u$. Having shown this, we get
\begin{align*}
&p(xw_{\mathit{Child}(u_{1})},\cdots,xw_{\mathit{Child}(u_{n(u)})},x_{\mathit{Child}(u)^c},w_{\mathit{Sub}(u)\setminus\mathit{Child}(u)})|x_u) \\
&\hspace{-.3 cm}=p(x_{\mathit{Child}(u_{1})},\cdots,x_{\mathit{Child}(u_{n(u)})},x_{\mathit{Child}(u)^c}|x_u)\cdot p(w_{\mathit{Child}(u_{1})},\cdots,w_{\mathit{Child}(u_{n(u)})},w_{\mathit{Sub}(u)\setminus\mathit{Child}(u)})|x_{\mathcal{V}}) \\
&\hspace{-.3 cm}\stackrel{(a)}{=}(\prod \limits_{i=1}^{n(u)} p(x_{\mathit{Child}(u_{i})}|x_u))\cdot p(x_{\mathit{Child}(u)^c}|x_u) \cdot p(w_{\mathit{Child}(u_{1})},\cdots,w_{\mathit{Child}(u_{n(u)})},w_{\mathit{Sub}(u)\setminus\mathit{Child}(u)})|x_{\mathcal{V}}) \\
&\hspace{-.3 cm}\stackrel{(b)}{=}(\prod \limits_{i=1}^{n(u)} p(x_{\mathit{Child}(u_{i})}|x_u))\cdot p(x_{\mathit{Child}(u)^c}|x_u) \cdot (\prod\limits_{i=1}^{n(u)}p(w_{\mathit{Child}(u_{i})}|x_u,x_{\mathit{Child}(u_{i})}))\cdot  p(w_{\mathit{Sub}(u)\setminus\mathit{Child}(u)}|x_u,x_{\mathit{Child}(u)^c}) \\
&\hspace{-.3 cm}=(\prod \limits_{i=1}^{n(u)} p(xw_{\mathit{Child}(u_{i})}|x_u))\cdot p(x_{\mathit{Child}(u)^c},w_{\mathit{Sub}(u)\setminus\mathit{Child}(u)}|x_u)
\end{align*}
where $(a)$ follows from the Markov property (Definition \ref{def:MarkovProperty}) and where $(b)$ follows from a repeated use of   \eqref{eq:claimIndMarkov1} for the vertices in $${\mathit{In}}(u)\cup \{{\mathit{Sub}}(u) \setminus {\mathit{Child}}(u)\}$$ with respect to their ordering values. This completes the proof Claim~\ref{claim:treeIndependence}.

We now establish that \eqref{eq:claimIndMarkov1} holds for any $u$ by induction. For $u=1$, 
 the Markov chain \eqref{eq:claimIndMarkov1} reduces to
 \begin{align}
W_{u}-X_{u}-X_{\mathit{Child}(u)^c}
\end{align}
which is the same as the Markov chain \eqref{eq:treeAchieveMarkovs} for $u=1$. 

Assuming \eqref{eq:claimIndMarkov1} holds for $u=i$, $1 \leq i \leq k-1$, we show that the Markov chain \eqref{eq:claimIndMarkov1} holds for $u=k$.  

Write $\mathit{Sub}(u)\setminus\mathit{Child}(u)$ as
$$\mathit{Sub}(u)\setminus\mathit{Child}(u)=\mathit{Child}(v_1)\cup \mathit{Child}(v_2)\cup \cdots \cup \mathit{Child}(v_l) $$
with $$\mathit{Child}(v_i)\cap\mathit{Child}(v_j)=\varnothing\qquad 1 \leq i < j \leq l$$ where $l$ depends on $u$ and the ordering. We then have
\begin{align*}
p&(w_{\mathit{Child}(u)},w_{\mathit{Sub}(u)\setminus\mathit{Child}(u)}|x_{\mathcal{V}})\\
&=p(w_{\mathit{Child}(u)\setminus\{u\}},w_{\mathit{Sub}(u)\setminus\mathit{Child}(u)}|x_{\mathcal{V}}) \cdot p(w_u|w_{\mathit{Child}(u)\setminus\{u\}},w_{\mathit{Sub}(u)\setminus\mathit{Child}(u)},x_{\mathcal{V}})\\
&\stackrel{(a)}{=}p(w_{\mathit{Child}(u)\setminus\{u\}},w_{\mathit{Sub}(u)\setminus\mathit{Child}(u)}|x_{\mathcal{V}}) \cdot p(w_u|w_{\mathit{Child}(u)\setminus\{u\}},x_{\mathit{Child}(u)})\\
&=p(w_{\mathit{Child}(u_1)},\cdots,w_{\mathit{Child}(u_{n(u)})},w_{\mathit{Child}(v_1)},\cdots,w_{\mathit{Child}(v_{l})}|x_{\mathcal{V}}) \cdot p(w_u|w_{\mathit{Child}(u)\setminus\{u\}},x_{\mathit{Child}(u)})\\
&\stackrel{(b)}{=}p(w_{\mathit{Child}(u_1)},\cdots,w_{\mathit{Child}(u_{n(u)})}|x_{\mathit{Child}(u)})\cdot p(w_{\mathit{Child}(v_1)},\cdots,w_{\mathit{Child}(v_{l})}|x_{\mathcal{V}}) \cdot p(w_u|w_{\mathit{Child}(u)\setminus\{u\}},x_{\mathit{Child}(u)})\\
&=p(w_{\mathit{Child}(u)}|x_{\mathit{Child}(u)})\cdot p(w_{\mathit{Sub}(u)\setminus\mathit{Child}(u)}|x_{\mathcal{V}}) 
\end{align*}
which implies that
\begin{align*}
p&(w_{\mathit{Child}(u)}|x_{\mathcal{V}},w_{\mathit{Sub}(u)\setminus\mathit{Child}(u)})=p(w_{\mathit{Child}(u)}|x_{\mathit{Child}(u)})
\end{align*}
which shows the validity of the Markov chain \eqref{eq:claimIndMarkov1} for $u=k$.

Equality $(a)$ holds because of \eqref{eq:treeAchieveMarkovs} and equality $(b)$ follows from a repeated use of   \eqref{eq:claimIndMarkov1} for vertices $u_1$, $\cdots$, $u_{n(u)}$, $v_1$, $\cdots$, $v_l$ with respect to their ordering values---these Markov chains hold by the induction assumption. This completes the proof of Claim~\ref{claim:treeIndependence}.

\subsection{Proof of Claim~\ref{claim:treeGraphEntropy}}
\label{subSec:treeGraphEntropy}
Suppose that 
\begin{align}
G_{X_{\mathit{Child}(k)}|X_{\mathit{Sup}(k)},W_{\mathit{Roots}(k)}} = G_{X_{{\mathit{Child}(k)}}|X_{\mathit{Child}(k)^c}}\label{eq:Claim2Graphs}
\end{align}
holds. Then, we get 
\begin{align*}
H(&G_{X_{\mathit{Child}(k)}|X_{\mathit{Sup}(k)},W_{\mathit{Roots}(k)}})\\
&= \min \limits_{\substack{ V -
X_{\mathit{Child}(k)}- (X_{\mathit{Sup}(k)},W_{\mathit{Roots}(k)})\\ X_{\mathit{Child}(k)} \in V \in \Gamma(G_{X_{\mathit{Child}(k)}|X_{\mathit{Sup}(k)},W_{\mathit{Roots}(k)}})}} I(X_{\mathit{Child}(k)};V|X_{\mathit{Sup}(k)},W_{\mathit{Roots}(k)})\\
&=\min \limits_{\substack{ V -
X_{\mathit{Child}(k)}- (X_{\mathit{Sup}(k)},W_{\mathit{Roots}(k)})\\ X_{\mathit{Child}(k)} \in V \in \Gamma(G_{X_{\mathit{Child}(k)}|X_{\mathit{Sup}(k)},W_{\mathit{Roots}(k)}})}} H(V|X_{ \mathit{Sup}(k)},W_{\mathit{Roots}(k)})-H(V|X_{\mathit{Child}(k)})\\
&\stackrel{(a)}{=}\min \limits_{\substack{ V -
X_{\mathit{Child}(k)}- (X_{\mathit{Sup}(k)},W_{\mathit{Roots}(k)})\\ X_{\mathit{Child}(k)} \in V \in \Gamma(G_{X_{\mathit{Child}(k)}|X_{\mathit{Sup}(k)},W_{\mathit{Roots}(k)}})}} H(V|X_{\mathit{Sup}(k)})-H(V|X_{\mathit{Child}(k)})\\
&=\min \limits_{\substack{ V -
X_{\mathit{Child}(k)}- X_{\mathit{Sup}(k)}\\ X_{\mathit{Child}(k)} \in V \in \Gamma(G_{X_{\mathit{Child}(k)}|X_{\mathit{Sup}(k)},W_{\mathit{Roots}(k)}})}} I(X_{\mathit{Child}(k)};V|X_{\mathit{Sup}(k)})\\
&\stackrel{(b)}{=}\min \limits_{\substack{ V -
X_{\mathit{Child}(k)}- X_{\mathit{Child}(k)^c}\\ X_{\mathit{Child}(k)} \in V \in \Gamma(G_{X_{\mathit{Child}(k)}|X_{\mathit{Sup}(k)},W_{\mathit{Roots}(k)}})}} I(X_{\mathit{Child}(k)};V|X_{\mathit{Child}(k)^c})\\
&\stackrel{(c)}{=}\min \limits_{\substack{ V -
X_{\mathit{Child}(k)}- X_{\mathit{Child}(k)^c}\\ X_{\mathit{Child}(k)} \in V \in \Gamma(G_{X_{\mathit{Child}(k)}|X_{\mathit{Child}(k)^c }})}} I(X_{\mathit{Child}(k)};V|X_{\mathit{Child}(k)^c})\\
&=H(G_{X_{\mathit{Child}(k)}|X_{\mathit{Child}(k)^c}}).
\end{align*}
Equality $(a)$ holds because the Markov chains $$V -
X_{\mathit{Child}(k)}- (X_{\mathit{Sup}(k)},W_{\mathit{Roots}(k)})$$ and $$X_{\mathit{Child}(k)}- X_{ \mathit{Sup}(k)}-W_{\mathit{Roots}(k)},$$ which are due to Claim~\ref{claim:treeIndependence}, imply the Markov chain 
$V - X_{\mathit{Sup}(k)}-W_{\mathit{Roots}(k)}$. Equality $(b)$ holds by the Markov property (Definition~\ref{def:MarkovProperty}). Finally $(c)$ holds by \eqref{eq:Claim2Graphs}.

We now show the graph equality \eqref{eq:Claim2Graphs}. First observe that the vertex sets in these two graphs are the same and equal to $\mathcal{X}_{\mathit{Child}(k)}$. It remains to show that any two vertices $x_{\mathit{Child}(k)}$ and $x'_{\mathit{Child}(k)}$ in $G_{X_{\mathit{Child}(k)}|X_{\mathit{Sup}(k)},W_{\mathit{Roots}(k)}}$ are connected if and only if they are connected in $G_{X_{{\mathit{Child}(k)}}|X_{\mathit{Child}(k)^c}}$.

\begin{itemize}
\item Suppose that $x_{\mathit{Child}(k)}$ and $x'_{\mathit{Child}(k)}$ are connected in $G_{X_{\mathit{Child}(k)}|X_{\mathit{Sup}(k)},W_{\mathit{Roots}(k)}}$. By Definition~\ref{def:condCharGraph} this means that there exist $x_{\mathit{Child}(k)^c }$ and $w_{\mathit{Roots}(k)}$ such that
\begin{align}
p(x_{\mathit{Child}(k)},x_{\mathit{Child}(k)^c},w_{\mathit{Roots}(k)})\cdot &p(x'_{\mathit{Child}(k)},x_{\mathit{Child}(k)^c},w_{\mathit{Roots}(k)})>0 \label{eq:Claim2Case1P} \\
f(x_{\mathit{Child}(k)},x_{\mathit{Child}(k)^c})&\neq f(x'_{\mathit{Child}(k)},x_{\mathit{Child}(k)^c}). \label{eq:Claim2Case1f}
\end{align}
Inequality \eqref{eq:Claim2Case1P} yields $$p(x_{\mathit{Child}(k)},x_{\mathit{Child}(k)^c})\cdot p(x'_{\mathit{Child}(k)},x_{\mathit{Child}(k)^c})>0$$ which, together with \eqref{eq:Claim2Case1f}, implies that $x_{\mathit{Child}(k)}$ and $x'_{\mathit{Child}(k)}$ are connected in $G_{X_{\mathit{Child}(k)}|X_{\mathit{Child}(k)^c}}$.
\item Suppose that $x_{\mathit{Child}(k)}$ and $x'_{\mathit{Child}(k)}$ are connected in $G_{X_{\mathit{Child}(k)}|X_{\mathit{Child}(k)^c}}$. Due to Definition~\ref{def:condCharGraph}, this means that there exists $x_{\mathit{Child}(k)^c }$  such that
\begin{align}
p(x_{\mathit{Child}(k)},x_{\mathit{Child}(k)^c})\cdot &p(x'_{\mathit{Child}(k)},x_{\mathit{Child}(k)^c})>0 \label{eq:Claim2Case2P} \\
f(x_{\mathit{Child}(k)},x_{\mathit{Child}(k)^c})&\neq f(x'_{\mathit{Child}(k)},x_{\mathit{Child}(k)^c}). \label{eq:Claim2Case2f}
\end{align}
Inequality \eqref{eq:Claim2Case2P} and the Markov chain
$$X_{\mathit{Child}(k)}-X_{\mathit{Child}(k)^c}-W_{\mathit{Roots}(k)}$$
obtained from Claim~\ref{claim:treeIndependence} imply that there exists $w_{\mathit{Roots}(k)}$ such that $$p(x_{\mathit{Child}(k)},x_{\mathit{Child}(k)^c},w_{\mathit{Roots}(k)})\cdot p(x'_{\mathit{Child}(k)},x_{\mathit{Child}(k)^c},w_{\mathit{Roots}(k)})>0.$$ This together with \eqref{eq:Claim2Case2f} implies that $x_{\mathit{Child}(k)}$ and $x'_{\mathit{Child}(k)}$ are connected in $$G_{X_{\mathit{Child}(k)}|X_{\mathit{Sup}(k)},W_{\mathit{Roots}(k)}}.$$
\end{itemize}

\subsection{Proof of Claim~\ref{claim:WuisinWAina}}
\label{subSec:WuisinWAina}
The Markov chain follows from
\begin{align}
p(w_k^{'}|w_{\mathit{In}(k)}^{*},x_k,&x_{\mathit{Child}(k)^c},w^{*}_{\mathit{Sub}(k)\setminus \mathit{Child}(k)})\nonumber \\
&=\sum \limits_{x_{\mathit{Child}(k) \setminus \{k\}}} p(w_k^{'}|w_{\mathit{In}(k)}^{*},x_{\mathit{Child}(k)},x_{\mathit{Child}(k)^c},w^{*}_{\mathit{Sub}(k)\setminus \mathit{Child}(k)})\nonumber \\
& \hspace{1.85 cm} \cdot p(x_{\mathit{Child}(k) \setminus \{k\}}|w_{\mathit{In}(k)}^{*},x_k,x_{\mathit{Child}(k)^c},w^{*}_{\mathit{Sub}(k)\setminus \mathit{Child}(k)}) \nonumber 
\\ 
&\stackrel{(a)}{=}\sum \limits_{x_{\mathit{Child}(k) \setminus \{k\}}} p(w_k^{'}|w_{\mathit{In}(k)}^{*},x_{\mathit{Child}(k)})\cdot p(x_{\mathit{Child}(k) \setminus \{k\}}|w_{\mathit{In}(k)}^{*},x_k,x_{\mathit{Child}(k)^c},w^{*}_{\mathit{Sub}(k)\setminus \mathit{Child}(k)}) \nonumber  
\\ 
&\stackrel{(b)}{=}\sum \limits_{x_{\mathit{Child}(k) \setminus \{k\}}} p(w_k^{'}|w_{\mathit{In}(k)}^{*},x_{\mathit{Child}(k)})\cdot  p(x_{\mathit{Child}(k) \setminus \{k\}}|w_{\mathit{In}(k)}^{*},x_k) \nonumber \\
&=p(w_k^{'}|w_{\mathit{In}(k)}^{*},x_k), \nonumber
\end{align}
where $(a)$ holds because of the Markov chain
\begin{align*}
&W^{'}_k-X_{\mathit{Child}(k)}-(X_{\mathit{Child}(k)^c},W^{*}_{\mathit{Sub}(k)})
\end{align*}
which can be deduced from Definition~\eqref{Milad2}. Equality $(b)$ follows from the Markov chains \eqref{eq:treeAchieveMarkovs} and Claim~\ref{claim:treeIndependence} applied to the vertices $\mathit{Sub}(k)$. \subsection{Proof of Claim~\ref{claim:treeEquiv}}
\label{subSec:treeEquiv}
Suppose $(w_{k_{1}},\cdots,w_{k_{n(k)}},x_k) \in \mathcal{B}_{w}$ and $p(x_{\mathit{Child}(k_{i})},w_{k_{i}})>0$, $1 \leq  i \leq n(k)$. The first term together with the definition of $\mathcal{B}_{w}$ implies that there exists $x'_{\mathit{Child}(k)\setminus\{k\}} \in \mathcal{X}_{\mathit{Child}(k)\setminus\{k\}}$ such that $p(x'_{\mathit{Child}(k_{i})},w_{k_{i}})>0$, $1 \leq  i \leq n(k)$, and $(x'_{\mathit{Child}(k_{1})},\cdots,x'_{\mathit{Child}(k_{n(k)})},x_k) \in w$. 
 
For any $x_{\mathit{Child}(k)^c} \in \mathcal{X}_{\mathit{Child}(k)^c}$ and $w_{\mathit{Roots}(k)}\in W^*_{\mathit{Roots}(k)}$ such that
\begin{align}
&p(x_{\mathit{Child}(k_{1})},\cdots,x_{\mathit{Child}(k_{n(k)})},x_k,x_{\mathit{Child}(k)^c},w_{\mathit{Roots}(k)})\nonumber \\
&\hspace{5 cm}\cdot p(x'_{\mathit{Child}(k_{1})},\cdots,x'_{\mathit{Child}(k_{n(k)})},x_k,x_{\mathit{Child}(k)^c},w_{\mathit{Roots}(k)})>0,\label{claim3:xIuc}
\end{align}
we have 
\begin{align}
f(x'_{\mathit{Child}(k_{1})},\cdots,x'_{\mathit{Child}(k_{n(k)})},&x_k,x_{\mathit{Child}(k)^c})\nonumber \\
&\stackrel{(a)}{=}f(x'_{\mathit{Child}(k_{1})},\cdots,x'_{\mathit{Child}(k_{n(k)-1})},x_{\mathit{Child}(k_{n(k)})},x_k,x_{\mathit{Child}(k)^c})\nonumber\\
&=f(x'_{\mathit{Child}(k_{1})},\cdots,x'_{\mathit{Child}(k_{n(k)-2})},x_{\mathit{Child}(k_{n(k)-1})},x_{\mathit{Child}(k_{n(k)})},x_k,x_{\mathit{Child}(k)^c})\nonumber\\
&\hspace{3 cm}\cdots \nonumber \\
&=f(x_{\mathit{Child}(k_{1})},\cdots,x_{\mathit{Child}(k_{n(k)})},x_k,x_{\mathit{Child}(k)^c}). \label{claim3:equalFunction1}
\end{align} 
 We justify equality $(a)$---the other equalities can be deduced similarly. Inequality \eqref{claim3:xIuc} yields
\begin{align*}
p(x_{\mathit{Child}(k_{1})},\cdots,x_{\mathit{Child}(k_{n(k)})},x_k,x_{\mathit{Child}(k)^c}) \cdot p(x'_{\mathit{Child}(k_{1})},\cdots,x'_{\mathit{Child}(k_{n(k)})},x_k,x_{\mathit{Child}(k)^c})>0.
\end{align*}
 Due to the Markov property (Definition~\ref{def:MarkovProperty}), the above inequality can be re-written as
\begin{align*}
p(x_k)\cdot \prod \limits_{i} p(x_{\mathit{Child}(k_{i)})}|x_k) \cdot \prod \limits_{i} p(x'_{\mathit{Child}(k_{i)})}|x_k) \cdot p(x_{\mathit{Child}(k)^c}|x_k)>0
\end{align*}
which implies
\begin{align}\label{condpr}
p(x_k)\cdot p(x_{\mathit{Child}(k_{n(k))})}|x_k) \cdot \prod \limits_{i} p(x'_{\mathit{Child}(k_{i)})}|x_k) \cdot p(x_{\mathit{Child}(k)^c}|x_k)>0.
\end{align}
Using the Markov property, \eqref{condpr} can be re-written as
\begin{align}\label{condpr2}
&p(x'_{\mathit{Child}(k_{1})},\cdots,x'_{\mathit{Child}(k_{n(k)-1})},x_{\mathit{Child}(k_{n(k)})},x_k,x_{\mathit{Child}(k)^c}) \nonumber \\
&\hspace{5 cm}\cdot p(x'_{\mathit{Child}(k_{1})},\cdots,x'_{\mathit{Child}(k_{n(k)})},x_k,x_{\mathit{Child}(k)^c})>0.
\end{align} 
By combining \eqref{condpr2},  the fact that $$p(x_{\mathit{Child}(k_{n(k)})},w_{k_{n(k)}})\cdot p(x'_{\mathit{Child}(k_{n(k)})},w_{k_{n(k)}})>0,$$ and the Markov chain
$$W_{k_{n(k)}}-X_{\mathit{Child}(k_{n(k)})}-X_{\mathit{Child}(k_{n(k)})^c}$$
deduced from Claim~\ref{claim:treeIndependence}, we get
 \begin{align}\label{condpr3}
&p(x'_{\mathit{Child}(k_{1})},\cdots,x'_{\mathit{Child}(k_{n(k)-1})},x_{\mathit{Child}(k_{n(k)})},x_k,x_{\mathit{Child}(k)^c},w_{k_{n(k)}}) \nonumber \\
&\hspace{5 cm}\cdot p(x'_{\mathit{Child}(k_{1})},\cdots,x'_{\mathit{Child}(k_{n(k)})},x_k,x_{\mathit{Child}(k)^c},w_{k_{n(k)}})>0.
\end{align}
Inequality \eqref{condpr3} and the Markov chain
$$W_{\mathit{Roots}(k_{n(k)})}-X_{\mathit{Child}(k_{n(k)})^c}-(W_{k_{n(k)}},X_{\mathit{Child}(k_{n(k)})})$$
deduced from Claim~\ref{claim:treeIndependence}, imply that there exists $w_{\mathit{Roots}(k_{n(k)})}\in W^*_{\mathit{Roots}(k_{n(k)})}$ such that 
\begin{align*}
p(x'_{\mathit{Child}(k_{1})},\cdots,x'_{\mathit{Child}(k_{n(k)-1})},&x_{\mathit{Child}(k_{n(k)})},x_k,x_{\mathit{Child}(k)^c},w_{k_{n(k)}},w_{\mathit{Roots}(k_{n(k)})})\cdot \\
& p(x'_{\mathit{Child}(k_{1}),},\cdots,x'_{\mathit{Child}(k_{n(k)})},x_k,x_{\mathit{Child}(k)^c},w_{k_{n(k)}},w_{\mathit{Roots}(k_{n(k)})})>0.
\end{align*}From this inequality and  $$w_{k_{n(k)}} \in \Gamma(G_{X_{k_{n(k)}},W^*_{\mathit{In}(k_{n(k)})}|W^*_{\mathit{Roots}(k_{n(k)},O)},X_{\mathit{Sup}(k_{n(k)},O)}})$$ we get $$
f(x'_{\mathit{Child}(k_{1})},\cdots,x'_{\mathit{Child}(k_{n(k)})},x_k,x_{\mathit{Child}(k)^c})=f(x'_{\mathit{Child}(k_{1})},\cdots,x'_{\mathit{Child}(k_{n(k)-1})},x_{\mathit{Child}(k_{n(k)})},x_k,x_{\mathit{Child}(k)^c}).$$
This justifies equality $(a)$ in \eqref{claim3:equalFunction1}. 
 
From \eqref{claim3:xIuc} and \eqref{claim3:equalFunction1} the vertices $$(x_{\mathit{Child}(k_{1})},\cdots,x_{\mathit{Child}(k_{n(k)})},x_k)$$ and  $$(x'_{\mathit{Child}(k_{1})},\cdots,x'_{\mathit{Child}(k_{n(k)})},x_k)$$ are not connected in $G_{X_{\mathit{Child}(k)}|X_{\mathit{Sup}(k)},W^*_{\mathit{Roots}(k)}}$. From Claim~\ref{claim:treeCliques} stated thereafter (and proved in Appendix~\ref{subSec:treeCliques}) we deduce that any maximal independent set in $G_{X_{\mathit{Child}(k)}|X_{\mathit{Sup}(k)},W^*_{\mathit{Roots}(k)}}$ that includes $(x'_{\mathit{Child}(k_{1})},\cdots,x'_{\mathit{Child}(k_{n(k)})},x_k)$ should also include $(x_{\mathit{Child}(k_{1})},\cdots,x_{\mathit{Child}(k_{n(k)})},x_k)$. Hence we have $(x_{\mathit{Child}(k_{1})},\cdots,x_{\mathit{Child}(k_{n(k)})},x_k) \in w$. 

\begin{claim} \label{claim:treeCliques}
Suppose that $$(x_{\mathit{Child}(k)\setminus \{k\}},x_k), (x'_{\mathit{Child}(k)\setminus \{k\}},x_k), (x^{''}_{\mathit{Child}(k)}) \in \mathcal{X}_{\mathit{Child}(k)},$$ that $$(x_{\mathit{Child}(\mathit{Child}(k)\setminus \{k\})},x_k)\qquad \text{and}\qquad (x'_{\mathit{Child}(k)\setminus \{k\}},x_k)$$ are not connected in
$G_{X_{\mathit{Child}(k)}|X_{\mathit{Sup}(k)},W^*_{\mathit{Roots}(k)}}$, and that
$$p(x_{\mathit{Child}(k)\setminus \{k\}},x_k)\cdot p(x'_{\mathit{Child}(k)\setminus \{k\}},x_k) \cdot p(x^{''}_{\mathit{Child}(k)})>0.$$
Then, $(x_{\mathit{Child}(k)\setminus \{k\}},x_k)$ and $(x^{''}_{\mathit{Child}(k)})$ are connected in the graph $G_{X_{\mathit{Child}(k)}|X_{\mathit{Sup}(k)},W^*_{\mathit{Roots}(k)}}$ if and only if \linebreak$(x'_{\mathit{Child}(k)\setminus \{k\}},x_k)$ and $(x^{''}_{\mathit{Child}(k)})$ are connected.
\end{claim}

\subsection{Proof of Claim~\ref{claim:WuisinWAin}}
\label{subSec:WuisinWAin}
The distribution of $W^{*}_k$ and the fact that $w^{*}_k$ and $\mathcal{B}_{w^{'}_k}$ are in one-to-one correspondence guarantee that $W^*_k$ satisfies a. and c. We now show $W^{*}_k$ also satisfies b. 

\begin{itemize} 
\item $(W^{*}_{\mathit{In}(k)},X_k) \in W^*_k$: We show that $(w^{*}_{\mathit{In}(k)},x_k) \in \mathcal{B}_{w^{'}_k}$ assuming that $$p(W^*_k=\mathcal{B}_{w^{'}_k}|W^{*}_{\mathit{In}(k)}=w^{*}_{\mathit{In}(k)},X_k=x_k)>0.$$ By the definitions of $W^*_k$ and $W^{'}_k$ we have
\begin{align*}
p(W^*_k=\mathcal{B}_{w^{'}_k}|W^{*}_{\mathit{In}(k)}
=w^{*}_{\mathit{In}(k)},X_k=x_k)=p(W^{'}_k=w^{'}_k|W^{*}_{\mathit{In}(k)}=w^{*}_{\mathit{In}(k)},X_k=x_k)>0.
\end{align*}
Claim~\ref{claim:WuisinWAina} says that $$W_k^{'}-(X_k,W^{*}_{\mathit{In}(k)})-(X_{\mathit{Child}(k)^c },W^{*}_{\mathit{Sub}(k)\setminus \mathit{Child}(k)}),$$ 
and Claim~\ref{claim:treeIndependence} then implies that there exists $x_{\mathit{Child}(k)\setminus \{k\}}$ such that  $x_{\mathit{Child}(k)}  \in w_k^{'}$ and $$p(x_{\mathit{Child}(k_{i})},w^*_{k_{i}})>0, i \in \{1,2,\ldots,n(k)\}.$$ Hence, $(w^{*}_{\mathit{In}(k)},x_k) \in \mathcal{B}_{w^{'}_k}$.

\item $ W^*_k \in \Gamma^*(G_{W^{*}_{\mathit{In}(k)},X_k|X_{\mathit{Sup}(k)},W^{*}_{\mathit{Roots}(k)}})$: Consider  $w^*_k=\mathcal{B}_{w^{'}_k} \in \mathcal{W}^*$ and 
\begin{align}
p(w_{\mathit{In}(k)}(1),x_k,w^*_k)\cdot p(w_{\mathit{In}(k)}(2),x'_k,w^*_k)>0. \label{eq:treeClaimIndepSets}
\end{align}
 We now show that for any $x_{\mathit{Sub}(k)},x'_{\mathit{Sub}(k)} \in \mathcal{X}_{ \mathit{Sub}(k)}$,  $x_{\mathit{Sup}(k)} \in \mathcal{X}_{\mathit{Sup}(k)}$, and $w_{\mathit{Roots}(k)} \in \mathcal{W}^*_{\mathit{Roots}(k)}$ such that
\begin{align}
p(w_{\mathit{In}(k)}(1),x_{\mathit{Sub}(k)},&x_k,x_{\mathit{Sup}(k)},w_{\mathit{Roots}(k)}) \cdot p(w_{\mathit{In}(k)}(2),x'_{\mathit{Sub}(k)},x'_k,x_{\mathit{Sup}(k)},w_{\mathit{Roots}(k)})>0, \label{eq:treeEqui1}
\end{align}
we have
\begin{align}
f(x_{\mathit{Sub}(k)},x_k,x_{\mathit{Sup}(k)})=f(x'_{\mathit{Sub}(k)},x'_k,x_{\mathit{Sup}(k)}). \label{eq:treeEqui2}
\end{align}
Note that \eqref{eq:treeClaimIndepSets} and the distribution of $W^*_k$ imply that 
\begin{align}
p(w_{\mathit{In}(k)}(1),x_k,w^{'}_k)\cdot p(w_{\mathit{In}(k)}(2),x'_k,w^{'}_k)>0.  \nonumber
\end{align}
This, \eqref{eq:treeEqui1}, and the Markov chain
$$W_k^{'}-(X_k,W^{*}_{\mathit{In}(k)})-(X_{\mathit{Child}(k)^c},W^{*}_{\mathit{Sub}(k)\setminus \mathit{Child}(k)})$$
obtained from Claim~\ref{claim:WuisinWAina}, imply that
\begin{align}
p(w^{'}_k,x_{\mathit{Sub}(k)},x_k,x_{\mathit{Sup}(k)},w_{\mathit{Roots}(k)}) \cdot p(w^{'}_k,x'_{\mathit{Sub}(k)},x'_k,x_{\mathit{Sup}(k)},w_{\mathit{Roots}(k)})>0. \nonumber
\end{align}
From this inequality and the fact that $w^{'}_k \in \Gamma^*(G_{X_{\mathit{Child}(k)}|X_{\mathit{Sup}(k)},W^{*}_{\mathit{Roots}(k)}})$ we deduce \eqref{eq:treeEqui2}. We just showed that  $w^{*}_k$ is an independent set. We now show that it is maximal by way of contradiction.  

Let $w'$ be a maximal independent set in $$G_1\defeq G_{X_{\mathit{Child}(k)}|X_{\mathit{Sup}(k)},W^{*}_{\mathit{Roots}(k)}}$$
such that $$ w^{*}_k={\mathcal{B}}_{w'}.$$ 
 
Suppose that $w\defeq w^{*}_k$ is a subset of vertices that is not maximal in the graph 
$$G_2\defeq G_{W^{*}_{\mathit{In}(k)},X_k|X_{\mathit{Sup}(k)},W^{*}_{\mathit{Roots}(k)}}\,.$$
This means that $G_2$ contains a vertex $v\notin w$ that is not connected to any of the vertices in $w$. The fact that $v\notin w$ together with the definition of ${\mathcal{B}}_{w'}$ implies that there exists a vertex $q$ in $G_1$
 such that $q\notin w'$ and $p(q,v)>0$. Because of the latter and since $v$ is not connected to any of the vertices in $w$ we deduce that $q$ is not connected to any vertex in $w'$ from the definition of $G_2$ and Claim~\ref{claim:treeIndependence}. 
Finally, since $q\notin w'$ and since  $q$ is not connected to any vertex in $w'$, we deduce that the set of vertices
$$w'\cup \{q\}$$
is an independent set, a contradiction since $w'$ was supposed to be a maximal independent set. 
\end{itemize}

\subsection{Proof of Claim~\ref{claim:treeCliques}}
\label{subSec:treeCliques}
Suppose that $(x_{\mathit{Child}(k)\setminus \{k\}},x_k)$ and $(x^{''}_{\mathit{Child}(k)}) \in \mathcal{X}_{\mathit{Child}(k)}$ are connected in $G_{X_{\mathit{Child}(k)}|X_{\mathit{Sup}(k)},W^*_{\mathit{Roots}(k)}}$. This means that for some $x_{\mathit{Child}(k)^c} \in \mathcal{X}_{\mathit{Child}(k)^c}$ and $w_{\mathit{Roots}(k)}\in W^*_{\mathit{Roots}(k)}$ such that
\begin{align}
p(x_{\mathit{Child}(k)\setminus \{k\}},x_k,x_{\mathit{Child}(k)^c},w_{\mathit{Roots}(k)})\cdot p(x^{''}_{\mathit{Child}(k)},x_{\mathit{Child}(k)^c},w_{\mathit{Roots}(k)})>0,\label{eq:claim3x1yz}
\end{align}
we have 
\begin{align}
f(x_{\mathit{Child}(k)\setminus \{k\}},x_k,x_{\mathit{Child}(k)^c})\neq f(x^{''}_{\mathit{Child}(k)},x_{\mathit{Child}(k)^c}).  \label{eq:claim3fx1x'}
\end{align} 
 Note that \eqref{eq:claim3x1yz} implies
$$p(x_{\mathit{Child}(k)\setminus \{k\}},x_k,x_{\mathit{Child}(k)^c},w_{\mathit{Roots}(k)})>0$$
which, using Claim~\ref{claim:treeIndependence}, can be re-written as
$$p(x_{\mathit{Child}(k)\setminus \{k\}},x_k) \cdot p(x_{\mathit{Child}(k)^c},w_{\mathit{Roots}(k)}|x_k)>0.$$
This inequality and the claim's assumption that $p(x'_{\mathit{Child}(k)\setminus \{k\}},x_k)>0$  imply
$$p(x'_{\mathit{Child}(k)\setminus \{k\}},x_k) \cdot p(x_{\mathit{Child}(k)\setminus \{k\}},x_k) \cdot p(x_{\mathit{Child}(k)^c},w_{\mathit{Roots}(k)}|x_k)>0,$$
which, using Claim~\ref{claim:treeIndependence}, can be re-written as
\begin{align}
p(x_{\mathit{Child}(k)\setminus \{k\}},x_k,x_{\mathit{Child}(k)^c},w_{\mathit{Roots}(k)}) \cdot p(x'_{\mathit{Child}(k)\setminus \{k\}},x_k,x_{\mathit{Child}(k)^c},w_{\mathit{Roots}(k)})>0. \label{eq:claim8pxx'}
\end{align}
Using the claim's assumption that $(x_{\mathit{Child}(k)\setminus \{k\}},x_k)$ and $(x'_{\mathit{Child}(k)\setminus \{k\}},x_k)$ are not connected in
$$G_{X_{\mathit{Child}(k)}|X_{\mathit{Sup}(k)},W^*_{\mathit{Roots}(k)}},$$ we get 
\begin{align}
f(x_{\mathit{Child}(k)\setminus \{k\}},x_k,x_{\mathit{Child}(k)^c})=f(x'_{\mathit{Child}(k)\setminus \{k\}},x_k,x_{\mathit{Child}(k)^c}). \label{eq:claim8fxx'}
\end{align}
Now, \eqref{eq:claim3fx1x'} and \eqref{eq:claim8fxx'} yield
\begin{align}\label{ver1} f(x'_{\mathit{Child}(k)\setminus \{k\}},x_k,x_{\mathit{Child}(k)^c})\neq f(x^{''}_{\mathit{Child}(k)},x_{\mathit{Child}(k)^c}),
\end{align}
and \eqref{eq:claim3x1yz} and \eqref{eq:claim8pxx'} yield
\begin{align}\label{ver2}
p(x'_{\mathit{Child}(k)\setminus \{k\}},x_k,x_{\mathit{Child}(k)^c},w_{\mathit{Roots}(k)})\cdot p(x^{''}_{\mathit{Child}(k)},x_{\mathit{Child}(k)^c},w_{\mathit{Roots}(k)})>0.
\end{align}
From \eqref{ver1} and \eqref{ver2} we conclude that $(x'_{\mathit{Child}(k)\setminus \{k\}},x_k)$ and $(x^{''}_{\mathit{Child}(k)}) \in \mathcal{X}_{\mathit{Child}(k)}$ are also connected.

\subsection{Jointly Typical Sequences}\label{sect:typ}

\iffalse Let $x^n$ be a sequence with elements drawn from a finite alphabet $\mathcal{X}$. Define the empirical probability mass function of $x^n$ (or its type) as  
\begin{align}
\pi_{x^n}(x) \defeq \frac{|\{i:x_i=x\}|}{n} \hspace{.7 cm} x \in \mathcal{X}. \nonumber
\end{align}
Let $X\sim p(x)$. The set of $\varepsilon$-typical $n$-sequences is defined as  
\begin{align}
\mathcal{A}_{\varepsilon}^{(n)}(X)\defeq \{x^n:|\pi_{x^n}(x)-p(x)|\leq \varepsilon \cdot p(x) \text{, for all } x \in \mathcal{X}\}. \nonumber
\end{align}

Typical sequences satisfy the following properties.

\begin{lemma}[ {\cite[Lemmas~17,18,19]{OrliRoc01}}, {\cite[Page~26]{ElgaKim12}}] \label{lem:typicality}
The following claims hold:
\begin{itemize}
\item[a. ] Let $X^n\sim \prod_{i=1}^np_X(x_i)$.  Then, for $n$ large enough we have
$$P(X^n \in\mathcal{A}_{\varepsilon}^{(n)}(X))\geq 1- \delta(\varepsilon)$$
where $\delta(\varepsilon)\to 0$ as $\varepsilon \to 0$.
\item[b. ] For $n$ large enough we have
$$(1-\delta(\varepsilon))\cdot 2^{nH(X)(1-\varepsilon)} \leq |\mathcal{A}_{\varepsilon}^{(n)}(X)| \leq 2^{nH(X)(1+\varepsilon)}.$$
\item[c. ] Let $p(x^n)=\prod_{i=1}^np_X(x_i)$. Then, for each $x^n \in \mathcal{A}_{\varepsilon}^{(n)}(X)$,
\begin{itemize}
\item[i. ] $p_X(x_i)>0$  for all $1 \leq i \leq n$.
\item[ii. ] $2^{-nH(X)(1+\varepsilon)} \leq p(x^n)\leq 2^{-nH(X)(1-\varepsilon)}.$
\end{itemize}
\end{itemize}

\end{lemma}

In the same way jointly typicality is defined:
\fi

Let $(x^n,y^n)\in {\mathcal{X}}^n \times {\mathcal{Y}}^n $. Define the empirical probability mass function  of $(x^n,y^n)$ (or its type) as
\begin{align}
\pi_{x^n,y^n}(x,y) \defeq \frac{|\{i:(x_i,y_i)=(x,y)\}|}{n} \hspace{.5 cm}  (x,y)\in (\mathcal{X},\mathcal{Y}). \nonumber
\end{align}
Let $(X,Y)\sim p(x,y)$. The set of jointly $\varepsilon$-typical $n$-sequences is defined as  
\begin{align}
\mathcal{A}_{\varepsilon}^{(n)}(X,Y)\defeq \{(x^n,y^n):|\pi_{x^n,y^n}(x,y)-p(x,y)|\leq \varepsilon \cdot p(x,y)  \text{ for all } (x,y)\in (\mathcal{X},\mathcal{Y})\}. \nonumber
\end{align}
Also define the set of conditionally $\varepsilon$-typical $n$-sequences as
\begin{align}
\mathcal{A}_{\varepsilon}^{(n)}(Y|x^n)\defeq \{y^n:(x^n,y^n)\in \mathcal{A}_{\varepsilon}^{(n)}(X,Y)\}\,. \nonumber
\end{align}

Jointly typical sequences satisfy the following properties:

\begin{lemma}[{\cite[Corollary~2]{OrliRoc01}, \cite[Page~27]{ElgaKim12}}] \label{lem:jointlyTypicality}
For any $\varepsilon>0$ the following claims hold:
\begin{itemize}

\item[a. ] Let $(X^n,Y^n)\sim \prod_{i=1}^np_{X,Y}(x_i,y_i)$.  Then, for $n$ large enough we have
$$P((X^n,Y^n) \in\mathcal{A}_{\varepsilon}^{(n)}(X,Y))\geq 1- \delta(\varepsilon)$$
where $\delta(\varepsilon)\to 0$ as $\varepsilon \to 0$.
\item[b. ] For $n$ large enough we have
$(1-\delta(\varepsilon))2^{nH(X,Y)(1-\varepsilon)} \leq |\mathcal{A}_{\varepsilon}^{(n)}(X,Y)| \leq 2^{nH(X,Y)(1+\varepsilon)}.$
\item[c. ] Let $p(x^n,y^n)=\prod_{i=1}^np_{X,Y}(x_i,y_i)$. Then, for each $(x^n,y^n)\in  \mathcal{A}_{\varepsilon}^{(n)}(X,Y)$
\begin{itemize}
\item[i. ] $x^n \in \mathcal{A}_{\varepsilon}^{(n)}(X)$ and $y^n \in \mathcal{A}_{\varepsilon}^{(n)}(Y)$;
\item[ii. ] $p_{X,Y}(x_i,y_i)>0$ for all $1 \leq i \leq n$;
\item[iii. ] $2^{-nH(X,Y)(1+\varepsilon)} \leq p(x^n,y^n)\leq 2^{-nH(X,Y)(1-\varepsilon)} $;
\item[iv. ] 
$2^{-nH(X|Y)(1+\varepsilon)} \leq p(x^n|y^n)\leq 2^{-nH(X|Y)(1-\varepsilon)}$.
\end{itemize}
\end{itemize}
\end{lemma}
\begin{lemma}[Conditional Typicality Lemma, {\cite[Lemma~22]{OrliRoc01},  \cite[Page~27]{ElgaKim12}}] \label{lem:conditionalTypicalaity}  Fix $0<\varepsilon' < \varepsilon$, let $(X,Y)\sim p(x,y)$ and suppose that $x^n \in \mathcal{A}_{\varepsilon'}^{(n)}(X)$ and $Y^n\sim p(y^n|x^n)=\prod_{i=1}^np_{Y|X}(y_i|x_i)$. Then,  for $n$ large enough
$$P((x^n,Y^n) \in\mathcal{A}_{\varepsilon}^{(n)}(X,Y))\geq 1- \delta(\varepsilon,\varepsilon')$$
where $\lim_{\varepsilon\downarrow 0}\lim_{\varepsilon'\downarrow 0}\delta(\varepsilon,\varepsilon')=0$.
\end{lemma}

\begin{lemma}[Markov Lemma, {\cite[Lemma~23]{OrliRoc01}}]   \label{lem:markovTypicality} Let $X-Y-Z$ form a Markov chain. Suppose that $(x^n,y^n)\in \mathcal{A}_{\varepsilon'}^{(n)}(X,Y)$ and $Z^n\sim p(z^n|y^n)=\prod_{i=1}^np_{Z|Y}(z_i|y_i)$. Then, for $\varepsilon >\varepsilon'$ and $n$ large enough
$$P((x^n,y^n,Z^n) \in\mathcal{A}_{\varepsilon}^{(n)}(X,Y,Z))\geq 1- \delta(\varepsilon,\varepsilon').$$
\end{lemma}
\begin{lemma}[{\cite[Corollary~4]{OrliRoc01}}] \label{lem:independentTypicalaity} Let $p_{X,Y}(x,y)$ have marginal distributions $p_X(x)$ and $p_Y(y)$ and let $(X,Y)\sim p_{X,Y}(x,y)$. Let $(\mathbf{X}',\mathbf{Y}')\sim \prod_{i=1}^np_X(x'_i)\cdot p_Y(y'_i) $. Then, for $n$ large enough
\begin{align*}
(1- \delta(\varepsilon))\cdot 2^{-n(I(X;Y)+2\varepsilon H(Y))}\leq P((\mathbf{X}',\mathbf{Y}') \in\mathcal{A}_{\varepsilon}^{(n)}(X,Y))\leq 2^{-n(I(X;Y)-2\varepsilon H(Y))}\,.
\end{align*}
\end{lemma}
\begin{lemma}[Covering Lemma, {\cite[Lemma~3.3]{ElgaKim12}}]  \label{lem:covering} Let $(X,\hat{X})\sim p_{X,\hat{X}}(x,\hat{x})$. Let $X^n \sim \prod_{i=1}^n p_X(x_i)$ and $$\{\hat{X}^n(m), m \in \mathcal{B}\}\quad \text{with}\quad |\mathcal{B}|\geq 2^{nR}$$ be a set of random sequences independent of each other and of $X^n$, each distributed according to $\prod_{i=1}^n  p_{\hat{X}}(\hat{x}_i(m))$. Then, 
$$\lim \limits_{n \rightarrow \infty}P((X^n,\hat{X}^n(m)) \notin \mathcal{A}_{\varepsilon}^{(n)}(X,\hat{X}^n)\text{ for all }m \in \mathcal{B})=0,$$
if 
$$R>I(X;\hat{X})+\delta(\varepsilon).$$
\end{lemma}

\end{document}